\documentclass[aps,prb,reprint,superscriptaddress
]{revtex4-1}
\usepackage{graphicx}
\usepackage{amsmath,amsfonts,amssymb}
\usepackage{xcolor}
\usepackage{epstopdf}
\usepackage[retainplus]{siunitx}
\usepackage[version=4]{mhchem}
\usepackage{graphicx}
\usepackage{subfigure}
\usepackage{braket}
\usepackage{booktabs}
\usepackage{hyperref}
\hypersetup{
    colorlinks=true,
    linkcolor=blue,
    filecolor=magenta,      
    urlcolor=blue,
    citecolor=blue
    }

\definecolor{alizarin}{rgb}{0.82, 0.1, 0.26} 

\draft 

\begin{document}


 \title
  {Boron phosphide films by reactive sputtering: Searching for a p-type transparent conductor}



\author{Andrea Crovetto}
\email[]{Electronic mail: andrea.crovetto@helmholtz-berlin.de}
\affiliation{Materials Science Center, National Renewable Energy Laboratory, Golden, Colorado 80401, United States}
\affiliation{Department of Structure and Dynamics of Energy Materials, Helmholtz-Zentrum Berlin f\"ur Materialien und Energie GmbH, Berlin, Germany}

\author{Jesse M. Adamczyk}
\affiliation{Department of Physics, Colorado School of Mines, Golden, Colorado 80401, United States}

\author{Rekha R. Schnepf}
\affiliation{National Renewable Energy Laboratory, Golden, Colorado 80401, United States}
\affiliation{Department of Physics, Colorado School of Mines, Golden, Colorado 80401, United States}

\author{Craig L. Perkins}
\affiliation{Materials Science Center, National Renewable Energy Laboratory, Golden, Colorado 80401, United States}

\author{Hannes Hempel}
\affiliation{Department of Structure and Dynamics of Energy Materials, Helmholtz-Zentrum Berlin f\"ur Materialien und Energie GmbH, Berlin, Germany}

\author{Sage R. Bauers}
\affiliation{Materials Science Center, National Renewable Energy Laboratory, Golden, Colorado 80401, United States}

\author{Eric S. Toberer}
\affiliation{Department of Physics, Colorado School of Mines, Golden, Colorado 80401, United States}

\author{Adele C. Tamboli}
\affiliation{Materials Science Center, National Renewable Energy Laboratory, Golden, Colorado 80401, United States}
\affiliation{Department of Physics, Colorado School of Mines, Golden, Colorado 80401, United States}

\author{Thomas Unold}
\affiliation{Department of Structure and Dynamics of Energy Materials, Helmholtz-Zentrum Berlin f\"ur Materialien und Energie GmbH, Berlin, Germany}

\author{Andriy Zakutayev}
\email[]{Electronic mail: andriy.zakutayev@nrel.gov}
\affiliation{Materials Science Center, National Renewable Energy Laboratory, Golden, Colorado 80401, United States}

\begin{abstract}
With an indirect band gap in the visible and a direct band gap at a much higher energy, boron phosphide (BP) holds promise as an unconventional p-type transparent conductor. Previous experimental reports deal almost exclusively with epitaxial, nominally undoped BP films by chemical vapor deposition. High hole concentrations were often observed, but it is unclear if native defects alone can be responsible for it. Besides, the feasibility of alternative deposition techniques has not been clarified and optical characterization is generally lacking.
In this work, we demonstrate reactive sputtering of amorphous BP films, their partial crystallization in a P-containing annealing atmosphere, and extrinsic doping by C and Si. We obtain the highest hole concentration reported to date for p-type BP (\SI{5e20}{cm^{-3}}) using C doping under B-rich conditions. We also confirm that bipolar doping is possible in BP. An anneal temperature of at least \SI{1000}{\celsius} is necessary for crystallization and dopant activation. Hole mobilities are low and indirect optical transitions are much stronger than predicted by theory. Low crystalline quality probably plays a role in both cases. High figures of merit for transparent conductors might be achievable in extrinsically doped BP films with improved crystalline quality.

\end{abstract}

\pacs{}

\maketitle 

\section{Introduction}
Numerous III-V semiconductors such as GaAs, InP, GaN and related alloys are technologically mature materials. They are key components of optoelectronic devices such as light emitting diodes, photodetectors, lasers, and high-efficiency solar cells. Boron phosphide (BP) is a relatively under-investigated member of this family in spite of its ultra-high thermal conductivity,~\cite{Kang2017a} chemical inertness and hardness,~\cite{Stone1960} and prospects for bipolar doping.~\cite{Shohno1974}
Accordingly, BP has been proposed for several applications such as a corrosion- and heat-resistant coating,~\cite{Ginley1983,Kumashiro1990} photo- and electrocatalyst,~\cite{Shi2016,Mou2019} as well as for thermal management~\cite{Kang2017a} and extreme UV optics applications.~\cite{Huber2016}

More recently, BP was identified as a potential p-type transparent conductive material (TCM).~\cite{Varley2017} This is a particularly interesting prospect, because obtaining high p-type conductivity in optically transparent materials is still an unsolved challenge.~\cite{Fioretti2020,Willis2021} Unlike the case of other p-type TCM candidates, bipolar doping has been reported in BP by various authors.~\cite{Shohno1974,Takigawa1974,Iwami1975,Kumashiro1990,Varley2017} Thus, BP could be a unique example of a transparent material with both p-type and n-type doping capability.

BP crystallizes in the diamond-derived zincblende structure with tetrahedral coordination. Because the electronegativity difference between B and P is small, BP is a covalent solid and its band structure is closely related to that of Si and C in the diamond structure. The main difference is an intermediate size of the fundamental indirect band gap for BP ($\sim$ \SI{2.0}{eV})~\cite{Archer1964,Ha2020} mainly due to an intermediate bond length. Although this band gap corresponds to visible light, the direct band gap of BP is much wider and falls in the UV region ($\sim$ \SI{4.3}{eV}).~\cite{Schroten1998,Ha2020} The weakness of indirect transitions predicted for BP at room temperature~\cite{Ha2020} is the key factor that could make BP thin films sufficiently transparent for many TCM applications. For example, a 100~nm-thick BP film is expected to absorb negligible amounts of red-yellow light and less than 10\% of violet light according to first-principles calculations including electron-phonon coupling.~\cite{Ha2020}
With respect to electrical properties, BP has a highly disperse valence band produced by p orbitals, ensuring low hole effective masses ($0.35~m_\mathrm{e}$).~\cite{Varley2017} Unlike the case of diamond, the valence band maximum of BP lies at a relatively shallow energy with respect to the vacuum level. These are enabling features for high p-type dopability.~\cite{Zunger2003,Goyal2020}

\section{Open questions in BP research}
\subsection{Conductivity and transparency}
The highest conductivity reported for p-type BP is \SI{3600}{S/cm} for a nominally undoped single-crystalline film deposited by chemical vapor deposition (CVD) at \SI{1050}{\celsius} using B$_2$H$_6$ and PH$_3$ gas precursors in hydrogen, as reported by Shohno et al..~\cite{Shohno1974} Remarkably, this value is on par with the conductivities of the best n-type TCMs.~\cite{Gordon2000} The hole concentration and mobility were \SI{8e19}{cm^{-3}} and \SI{285}{cm^{2}/Vs} respectively.
However, intrinsic defects in BP are not expected to be effective dopants~\cite{Varley2017} so the origin of the high carrier concentration is unclear and requires further investigation. In addition, such a high conductivity has not been reproduced by others, with more commonly reported values being in the \SI{0.1}\textendash \SI{50}{S/cm} range for both n-type and p-type BP.~\cite{Stone1960,Wang1964,Chu1971,Iwami1975,Kato1977,Kumashiro1990}

A common figure of merit (FOM) for TCMs is the ratio between electrical conductivity and average absorption coefficient in the visible.~\cite{Gordon2000,Crovetto2020d} However, the absorption coefficient of BP has rarely been measured. Absorption coefficients of single crystals and single-crystalline films~\cite{Archer1964,Wang1964,Iwami1975} are generally below \SI{1e3}{cm^{-1}} at \SI{2.5}{eV}, in line with computational predictions. Yet, a more recent measurement on a BP thin film~\cite{Odawara2005} shows a substantially higher absorption coefficient, above \SI{1e4}{cm^{-1}} in most of the visible region. Furthermore, BP films on transparent substrates are usually described as orange,~\cite{Chu1971} red,~\cite{Takigawa1974} brown,~\cite{Iwami1975} or black,~\cite{Goossens1989} indicating that their optical transmission may not be sufficiently high.

The only case where electrical conductivity and absorption coefficient were measured on the same BP sample is the CVD-grown n-type single-crystalline BP film reported by Iwami et al.~\cite{Iwami1975} with a FOM of $\sim$ \SI{0.02}{\ohm^{-1}}. If we use the absorption coefficient from this study to estimate the FOM of the most conductive p-type BP films reported by Shohno et al.,~\cite{Shohno1974} we get a FOM of \SI{1.8}{\ohm^{-1}}. To contextualize these values, the highest FOM we are aware of for a p-type TCM of any kind is \SI{1.3}{\ohm^{-1}} for reactively sputtered CuI.~\cite{Yang2016a} Thus, BP deserves experimental scrutiny with focus on its applicability as a p-type TCM.

\subsection{Film growth}
BP is notoriously difficult to synthesize due to the simultaneous presence of a highly inert (B) and a volatile species (P).~\cite{Woo2016} High growth temperatures (typically above \SI{900}{\celsius}) are necessary to activate boron diffusion and obtain crystalline BP, but at these temperatures BP tends to decompose into boron-rich phosphides (such as B$_6$P) and gaseous phosphorus.~\cite{Kumashiro1997a,Shohno1974} For this reason, previous thin-film work on BP has given strong preference to CVD processes near atmospheric pressure where it is easier to prevent P losses by keeping a high P partial pressure during deposition. Attempts to grow BP in high vacuum by solid-source evaporation~\cite{Dalui2008} and gas-phase molecular beam epitaxy~\cite{Kumashiro1997a} resulted in amorphous films and insufficient P incorporation at elevated temperatures. Sputter deposition could be a viable alternative because it operates under intermediate pressure conditions. As a non-equilibrium plasma-assisted process, sputter deposition may enable lower temperature growth~\cite{Yoshioka2008} and higher dopant solubility in the host material~\cite{Bikowski2015} compared to thermal processes.
Besides, epitaxial single-crystalline CVD films grown above \SI{1000}{\celsius} are of limited applicability in the TCM industry, where cost-effective large-area coating of glass substrates or multilayer stacks are necessary. Magnetron sputtering, on the other hand, is the deposition method of choice for many n-type TCMs in the industry.~\cite{Morales-Masis2017,Szyszka2008,Seo2020} We are only aware of one previous attempt of sputter deposition of BP.~\cite{Jia2011} Optoelectronic characterization was not conducted and it is also unclear if the deposited films actually consisted of BP.

\begin{figure}[t!]
\centering%
\includegraphics[width=\columnwidth]{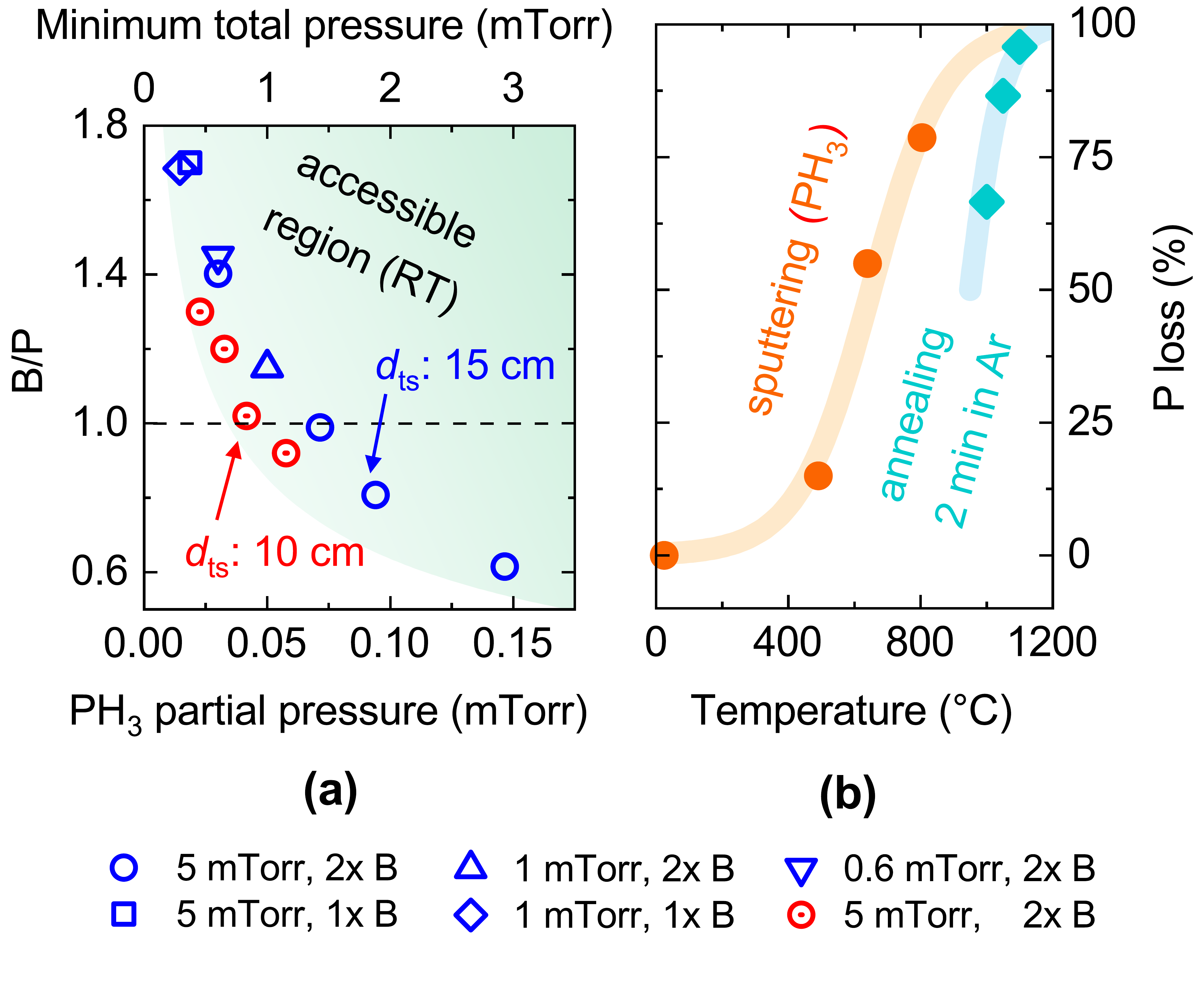}
\caption{Composition of B-P films. (a): Atomic B/P ratio of B-P films sputtered at room temperature versus PH$_3$ partial pressure. Datasets for two different target-substrate distances $d_\mathrm{ts}$ are shown. Different symbols refer to different total pressures and numbers of co-sputtering B targets ("2x B" meaning two targets).
Assuming that the PH$_3$ concentration cannot exceed 5\% of the Ar background, each  PH$_3$ partial pressure has a corresponding minimum total pressure at which the B/P region below the data points is not experimentally accessible. (b): Phosphorus loss in B-P films as a function of temperature. Circles refer to the loss of P during sputter deposition at fixed PH$_3$ partial pressure. Diamonds refer to the P loss of B-P films annealed at \SI{1}{bar} for 2~min under a continuous Ar flow. Lines are guides to the eye.}
\label{fig:composition}
\end{figure}

The goal of this work is to answer some of the important open questions elaborated in this section. First, we establish that amorphous BP films can be deposited by reactive sputtering and can be crystallized in a post-annealing step. Second, we find higher absorption coefficients than expected, possibly due to enhanced indirect transitions associated with imperfect crystallinity. Finally, we apply C doping to achieve bipolar conductivity and the highest hole concentration achieved so far in p-type BP.
Thus, we confirm that BP can be doped both n-type and p-type, and we propose that BP may reach a high TCM figure of merit by a combination of high crystalline quality and extrinsic doping.

\section{Results}
\subsection{One-step growth by reactive sputtering}
Smooth amorphous BP films can be grown by reactive sputtering of B targets in a PH$_3$/Ar atmosphere. The B/P ratio in the films, as measured by Auger electron spectroscopy (AES), can be continuously adjusted over a wide range at room temperature (Fig.~\ref{fig:composition}(a)). The main factor determining the B/P ratio at fixed temperature is the PH$_3$ partial pressure, with no significant role played by the total sputter pressure and the number of simultaneously sputtering B targets (one or two). A typical oxygen content in the bulk is 2~at.\% according to AES.

We infer from these observations that P incorporation in reactively sputtered BP neither occurs as a reaction at the target, nor as a direct reaction of PH$_3$ with elemental B at the substrate. A target reaction would result in an increasing B/P ratio with total pressure at constant PH$_3$ partial pressure, due to a lower fraction of reactive species bombarding the target. A reaction of PH$_3$ with elemental B at the substrate would lead to an increasing B/P ratio with increasing number of co-sputtering B targets. The most likely explanation is that PH$_3$ cracks into more reactive species in the plasma region in front of each target, with subsequent condensation of P at the substrate. The wide continuous range of B/P ratios that can be achieved in the films indicates that P condensation at the substrate results in formation of B-P bonds as well as P-P bonds. Since the flux of reactive P species from PH$_3$ cracking is less directional than the flux of B species ejected from the target, one could expect the films to be more B-rich at larger target-substrate distances, which we indeed observe in our data (Fig.~\ref{fig:composition}(a)). Due to the volatility of P at high temperatures (Fig.~\ref{fig:composition}(b)), a partial pressure of \SI{0.4}{mTorr} is necessary to obtain $\mathrm{B/P} = 1$ at a substrate temperature of \SI{800}{\celsius}. Therefore, sputtering BP at very high temperatures requires either a high PH$_3$ concentration in the sputter gas (beyond the 5\% limit dictated by our pre-diluted PH$_3$ bottle) or a high total sputter pressure.

\begin{figure}[t!]
\centering%
\includegraphics[width=\columnwidth]{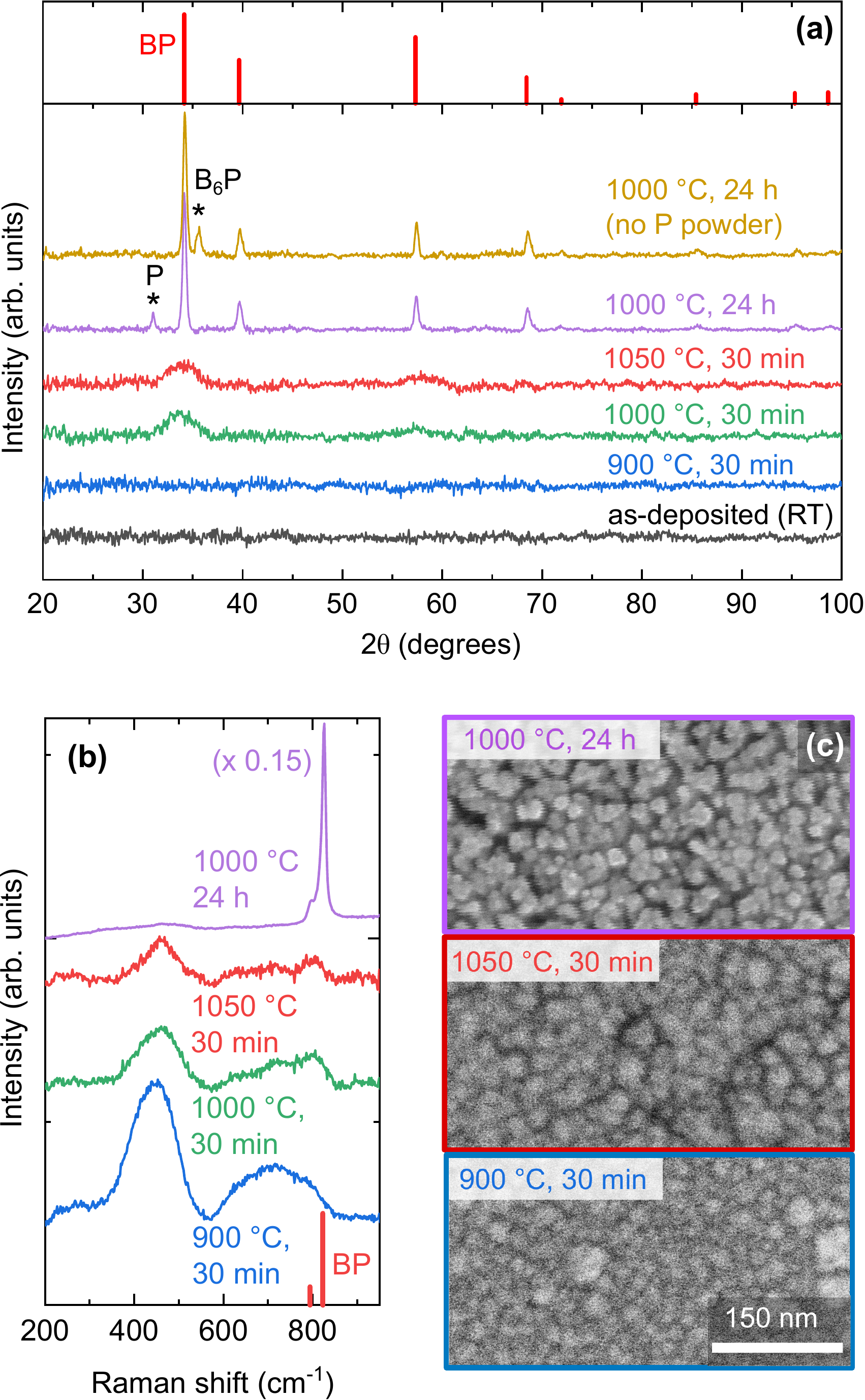}
\caption{Structure and morphology of BP films. (a): XRD patterns of various C-doped BP films annealed in sealed ampoules. Process conditions are indicated. The reference patterns for zincblende BP and rhombohedral B$_6$P are taken from the Inorganic Crystal Structure Database.~\cite{Xia1993,Morosin2008} (b): Raman spectra of four of the BP films shown in (a). The reference Raman spectrum of zincblende BP is shown.~\cite{Solozhenko2014} (c): SEM images of three of the BP films shown in (a).}
\label{fig:xrd_raman_sem}
\end{figure}

Regardless of substrate temperature (\SI{25}{\celsius} to \SI{800}{\celsius}) and stoichiometry ($0.5 < \mathrm{B/P} < 1.5$), the conductivity of all as-deposited BP films is too low to be measured on the plane of the substrate with a conventional four-point probe. An upper limit for their conductivity is about \SI{e-3}{S/cm}, indicating that native defects (such as B and P antisites) are ineffective dopants in this temperature range. 
To check if these amorphous films could be crystallized, we conducted rapid thermal annealing (RTA) at atmospheric pressure under a continuous Ar flow. Despite the higher total pressure, the lack of a P partial pressure in the system led to rapid P losses. After 2~min at \SI{1000}{\celsius}, the films have already lost 65\% of their original P content (Fig.~\ref{fig:composition}(b)). Very short anneals at \SI{1100}{\celsius} result in broad X-ray diffraction (XRD) peaks attributable to BP. However, the conductivity of all films after RTA is still below \SI{e-3}{S/cm}. From these initial experiments, we conclude that annealing in a P-containing atmosphere is necessary to prevent P losses and explore a wider process window. It also seems unlikely that intrinsic doping can lead to significant carrier concentrations in BP, so options for extrinsic dopants should be explored.

\subsection{Two-step growth by reactive sputtering and annealing with extrinsic doping}
To address the issues described above, we post-annealed BP films in sealed quartz ampoules in the presence of sacrificial red phosphorus powder. According to defect calculations, C and Si are the most attractive candidates for p-type doping of BP under B-rich conditions.~\cite{Varley2017} We found that C could be easily incorporated into BP by cleaning the ampoules with isopropyl alcohol before annealing (see Experimental Details). The C content of BP films annealed for 30~min is consistently in the 2.5\textendash 2.7~at.\% range, roughly independent of annealing temperature. This extrinsic dopant concentration is similar to the case of n-type TCMs such as ZnO:Al and In$_2$O$_3$:Sn.~\cite{Crovetto2016a,Utsumi2003} AES depth profiles indicate that the C impurities are homogeneously distributed throughout the depth of the films.

Raman spectra are very sensitive to C impurities in excess of their solubility in the BP matrix, due to the high scattering cross section of the $D$ and $G$  Raman bands in disordered carbon.~\cite{Nakamura1990} Although these Raman bands are already detected at 2.5~at.\% C in our samples (Fig.~\ref{fig:SIraman}, Supporting Information), this C impurity concentration is still sufficiently low not to have a significant effect on the optical properties of BP (Fig.~\ref{fig:SIabsorption}, Supporting Information). Based on an empirical relationship between the width of the $D$ band and the C crystallite size,~\cite{Nakamura1990} we estimate that non-substitutional C segregates in very small clusters around \SI{1}{nm} in size. Unless otherwise specified, the post-annealed BP films presented in the rest of the paper are C-doped. The annealing process also results in an increase of the O content in the bulk, from about 2\% to about 6\%. Two likely reasons are the incomplete removal of air when sealing the ampoules and the traces of isopropyl alcohol responsible for C doping. O impurities are unlikely to act as dopants in BP due to a large size mismatch with P.

With an annealing time of 30~min, a temperature of \SI{900}{\celsius} is still not sufficient to detect any crystalline BP by XRD (Fig.~\ref{fig:xrd_raman_sem}(a)). The corresponding Raman spectrum (Fig.~\ref{fig:xrd_raman_sem}(b)) exhibits two broad bands centered at $\sim$\SI{450}{cm^{-1}} and $\sim$\SI{700}{cm^{-1}}. These could be attributed to the acoustic and optical phonon bands of zincblende BP with significant broadening.~\cite{Ha2020} However, the Raman features of amorphous red phosphorus~\cite{Fasol1984} and amorphous B~\cite{Kuhlmann1994} are also located in these two spectral regions, so we cannot exclude the presence of the elemental forms of P and B. This film mainly consists of small ($\sim$\SI{20}{nm}) particles, with some larger particles that appear brighter in the scanning electron microscope (SEM) and might therefore consist of elemental P (Fig.~\ref{fig:xrd_raman_sem}(c)).

\begin{figure}[t!]
\centering%
\includegraphics[width=\columnwidth]{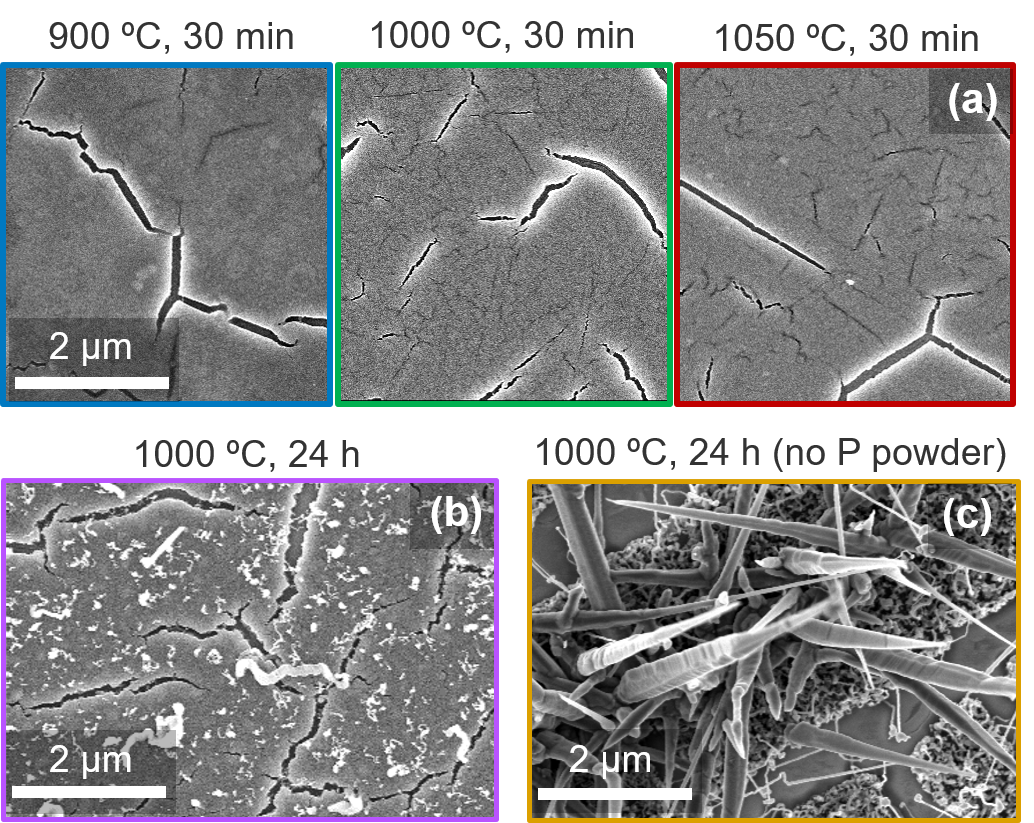}
\caption{Low-magnification SEM images of the BP films shown in Fig.~\ref{fig:xrd_raman_sem}(a). Annealing conditions are indicated. Films were annealed with (a,b) or without (c) sacrificial P powder The secondary phases present in the film annealed for 24~h (b) are mainly elemental P. When no sacrificial P powder is present (c) a whisker morphology develops.}
\label{fig:sem}
\end{figure}

When the annealing temperature is increased to \SI{1000}{\celsius}, a broad XRD peak with full width at half maximum (FWHM) of \SI{3.2}{\degree 2\theta} appears (Fig.~\ref{fig:xrd_raman_sem}(a)). It can be attributed to the (111) reflection in zincblende BP. A similar peak FWHM is obtained after annealing at \SI{1050}{\celsius}. Since these widths are much larger than instrumental broadening, and assuming that only size effects are responsible for broadening, we estimate a coherence volume of \SI{3}{nm} using the Scherrer equation. This crystallite size is much smaller than the average particle size ($\sim$\SI{30}{nm}) determined by SEM (Fig.~\ref{fig:xrd_raman_sem}(c)) implying that each particle is poorly crystallized, although some medium-range order exists in some regions. In agreement with these observations, the Raman spectra of the films annealed at \SI{1000}{\celsius} and \SI{1050}{\celsius} for 30~min resemble the phonon density of states (DOS) calculated for BP.~\cite{Ha2020} The peaks at \SI{804}{cm^{-1}} and \SI{716}{cm^{-1}} correspond to the two peaks in the optical phonon DOS. The peak at \SI{460}{cm^{-1}} has a corresponding peak in the acoustic phonon DOS. We will refer to these films as nanocrystalline BP in the rest of the article.

To obtain higher crystalline quality, it is necessary to increase the annealing time. After a 24~h anneal at \SI{1000}{\celsius}, all the main XRD peaks of zincblende BP can be detected with a FWHM of \SI{0.41}{\degree 2\theta} for the (111) reflection. Considering the Lorentzian broadening component of this peak and deconvolving instrumental broadening,~\cite{DeKeijser1982,Crovetto2016a} we estimate a crystallite size of \SI{34}{nm} using the Scherrer equation. SEM images of this film (Fig.~\ref{fig:xrd_raman_sem}(c)) show that the particle size is still in the \SI{30}{nm} diameter range as for the films annealed for a shorter time, but the particle boundaries are much sharper. We conclude that long annealing times enable full crystallization of the (originally amorphous) BP particles, although the particles still do not coalesce into larger grains.
The Raman spectrum of this sample (Fig.~\ref{fig:xrd_raman_sem}(b)) is similar to spectra reported for high-quality BP crystals and single-crystalline films.~\cite{Solozhenko2014,Woo2016,Padavala2018} The main peak at \SI{827}{cm^{-1}} and the secondary peak at \SI{797}{cm^{-1}} can be attributed to the Raman-active longitudinal (LO) and transverse (TO) optical phonon modes at the $\Gamma$ point, respectively.~\cite{Ha2020} Our 24~h-long anneals result in phase segregation on the film surface, clearly visible in Fig.~\ref{fig:sem}(b), and in a new XRD peak at \SI{31.0}{\degree 2\theta} (Fig.~\ref{fig:xrd_raman_sem}(a)). According to AES imaging (Fig.~\ref{fig:SIauger_map}, Supporting Information), these surface secondary phases mainly consist of elemental P. The new XRD peak is compatible with both fibrous red phosphorus~\cite{Ruck2005} and Hittorf's (violet) phosphorus.~\cite{Hittorf1865}

If red phosphorus powder is not present during annealing, part of BP decomposes into B$_6$P and gaseous P until the P partial pressure in the ampoule is sufficiently high to prevent further P evaporation. The occurrence of this reaction is confirmed by the XRD peak at \SI{35.5}{\degree 2\theta}, compatible with the highest-intensity reflection of B$_6$P. A peculiar ``whisker" morphology develops for some of the remaining BP in these films, with typical whisker diameters of roughly \SI{500}{nm} and lengths of a few $\mu$m (Fig.~\ref{fig:sem}(c). BP whisker growth has sometimes been observed in CVD BP films.~\cite{Motojima1980,Schroten1996,Chu1971} Finally, we note that all the post-annealed BP films presented here exhibit cracks (Fig.~\ref{fig:sem}), probably due to the thermal expansion coefficient mismatch between BP and the fused silica substrate.~\cite{Slack1975}

\subsection{Optical properties}
\begin{figure}[t!]
\centering%
\includegraphics[width=\columnwidth]{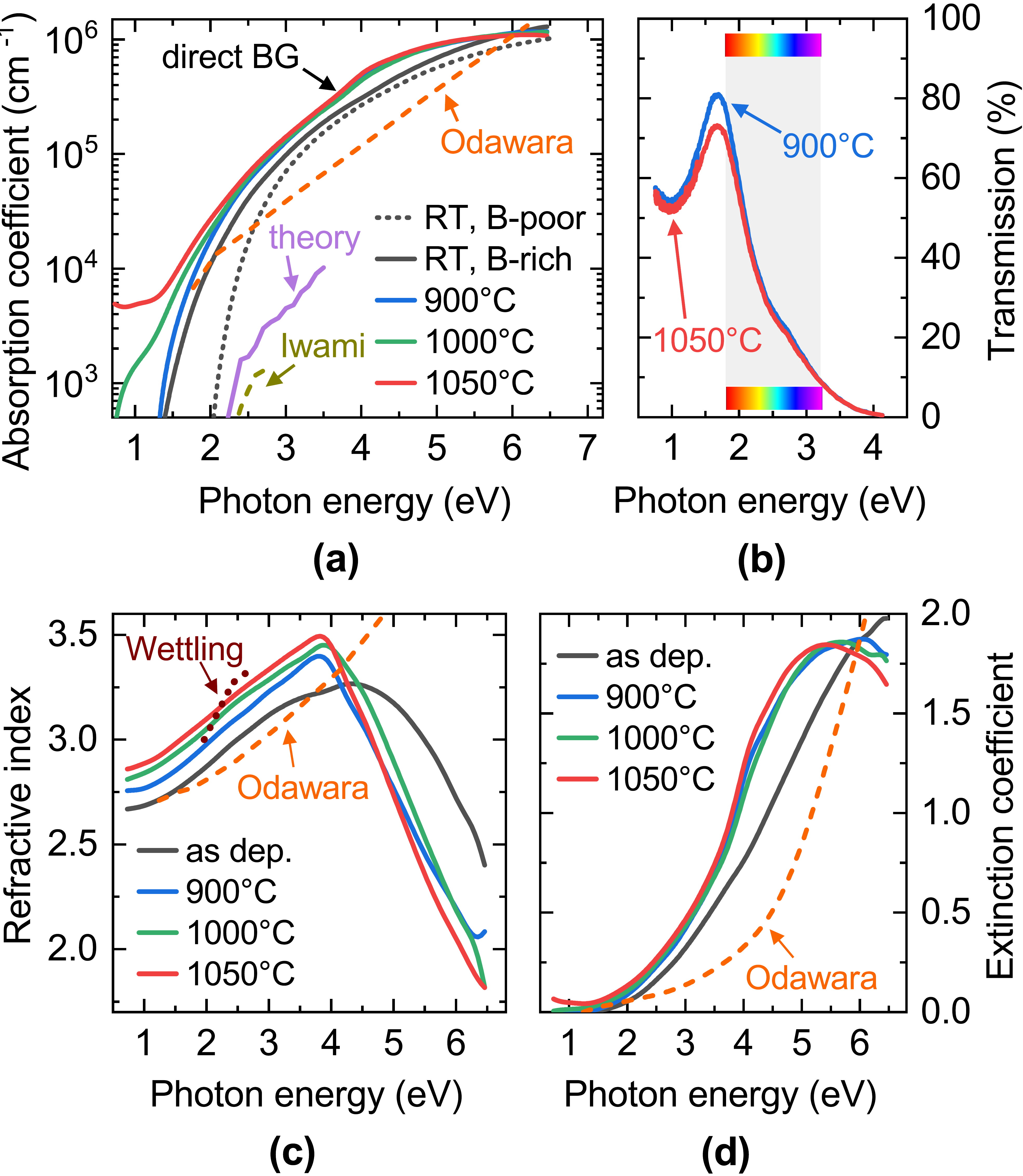}
\caption{Optical properties of as-deposited BP (black lines) and post-annealed BP (colored solid lines, 30~min annealing). (a) Absorption coefficient of B-rich films (B/P = 1.3, solid lines) and a B-poor as-deposited film (B/P = 0.6, dotted line). Post-annealed films were all obtained from the B-rich film. They reach a nearly stoichiometric composition after annealing. The computed absorption coefficient using a hybrid exchange correlation functional and electron-phonon coupling is shown.~\cite{Ha2020} Two previously measured absorption coefficients for BP films~\cite{Iwami1975,Odawara2005} are also shown. (b) Optical transmission of two BP films annealed at different temperatures. (c,d) Refractive index and extinction coefficient. The displayed films are the same as in (a). An additional experimental reference for the refractive index~\cite{Wettling1984} is included.}
\label{fig:optical}
\end{figure}

The absorption coefficient $\alpha$ of the as-deposited films depends on their stoichiometry, with B-rich films having lower band gaps than B-poor films (Fig.~\ref{fig:optical}(a)). In these amorphous films, $(\alpha h \nu)^{n}$ versus $h \nu$ plots are roughly linear for $ 1/3 < n < 1/2$ (Fig.~\ref{fig:SItauc}, Supporting Information), as usually found in amorphous semiconductors.~\cite{Davis1970} A similar spectral behavior and a decreasing band gap for increasing B content were found for amorphous B$_x$P films by CVD with $x > 3$.~\cite{Schroten1999} The band gaps of the B-rich and the B-poor films are estimated as \SI{1.3}{eV} and \SI{2.0}{eV}, respectively (Fig.~\ref{fig:SItauc}, Supporting Information). Note that the calculated indirect band gap of BP is 1.98~eV using hybrid exchange-correlation functionals.~\cite{Ha2020}

The absorption coefficient of the amorphous as-deposited films is much larger than the computational prediction for perfectly crystalline BP below the direct band gap at \SI{4}{eV}. This is expected because materials without long-range order do not require phonon participation in optical transitions. To achieve transparent conductivity in amorphous BP, a shift of the fundamental gap to higher photon energies with respect to the crystalline state would be advantageous to widen the transparency window. While this effect is very well known in amorphous Si,~\cite{Jellison1993a} we do not observe it in amorphous BP. In fact, a very recent molecular dynamics study predicts that amorphous BP should have an even lower band gap than crystalline BP.~\cite{Bolat2021} The reason is that even short-range order is often broken in amorphous BP, with abundant homoelement bonds (B-B and P-P) occurring in B-rich and B-poor clusters. As the band gap of amorphous boron is 1.0~eV,~\cite{Morita1975} the B-rich clusters are likely responsible for band gap narrowing in amorphous BP, as well as its further narrowing under B excess.~\cite{Schroten1999} Since this effect is unique to compound semiconductors, it may explain the discrepancy between the optical properties of amorphous BP and amorphous Si, in spite of their similar bonding and electronic structure in the crystalline state.

As shown in the next section and as observed by other workers,~\cite{Shohno1974,Kumashiro1990,Varley2017} B-rich conditions are necessary to achieve appreciable p-type conductivity in BP. Therefore, the remaining optical characterization shown in Fig.~\ref{fig:optical} focuses on B-rich films with B/P = 1.3. We limit our study to the nanocrystalline films annealed for 30~min, because longer anneals generate very rough films with secondary phases (Fig.~\ref{fig:sem}(b)), so their optical characterization is not reliable. The absorption coefficient of the post-annealed films is overall slightly larger than in the as-deposited film (Fig.~\ref{fig:optical}(a)). The post-annealed films also have a secondary absorption onset in the 4.0-4.5~eV range (Fig.~\ref{fig:SItauc}, Supporting Information) which is absent from the as-deposited films. This secondary onset is likely associated with the direct band gap of BP, predicted to lie at \SI{4.34}{eV} from computation~\cite{Ha2020} and measured at \SI{4.25}{eV} on crystalline BP.~\cite{Schroten1998} The primary absorption onset appears to shift to lower photon energies with increasing annealing temperature, although this effect might be an artifact often seen in rough films due to scattering and/or depolarization effects.~\cite{Swanepoel1984,Crovetto2018a,Crovetto2015} In any case, the visible absorption coefficient of our post-annealed films is over one order of magnitude larger than the absorption coefficient computed for pristine BP~\cite{Ha2020} and is also significantly larger than the experimental absorption coefficient determined for an epitaxial CVD film by ellipsometry (Fig.~\ref{fig:optical}(a)).~\cite{Odawara2005} As a consequence, the optical transmission of our $\sim$\SI{130}{nm}-thick films drops to very low values in the blue region of the visible (Fig.~\ref{fig:optical}(b)).

The refractive indices of our films match previous measurements on crystalline films~\cite{Wettling1984,Odawara2005} in the spectral region below the direct gap of BP (Fig.~\ref{fig:optical}(c)). At higher photon energies, the refractive indices decrease due to the inflection point in the extinction coefficient (Fig.~\ref{fig:optical}(d)). The slight increase in the refractive index and extinction coefficient with increasing annealing temperature is probably due to film densification upon annealing. The thickness of the uniform film layer defined in the ellipsometry model plus one half of the roughness layer decreases by $\sim$10\% between the as-deposited film and the film annealed at \SI{1050}{\celsius} (Table~\ref{tab:SIellipsometry_data}, Supporting Information). This matches the $\sim$10\% increase in refractive index seen in Fig.~\ref{fig:optical}(c)). By extrapolating the real part of the complex dielectric function to zero photon energy, we can estimate the high-frequency permittivity $\epsilon_\infty$ of BP, which is in the $7 < \epsilon_\infty < 8$ range for all films (Fig.~\ref{fig:SIdielectric_function}, Supporting Information). These values are consistent with the empirical band gap-permittivity relationship in III-V semiconductors.~\cite{Sirota1968} Since BP is strongly covalent, we expect the static permittivity $\epsilon_\mathrm{s}$ to be nearly equal to $\epsilon_\infty$.

\subsection{Electrical properties}

\subsubsection{Intrinsic and extrinsic dopants in BP}
The origin of the high carrier concentration reported for many BP crystals and films, either p-type~\cite{Stone1960,Wang1964,Chu1971,Shohno1974,Kumashiro1990} or n-type~\cite{Shohno1974,Iwami1975,Kato1977,Kumashiro1990} is unclear. First-principles calculations indicate that the lowest-energy acceptor level in BP is the B$_\mathrm{P}$ antisite.~\cite{Varley2017} However, this defect has a deep charge transition level (more than \SI{0.5}{eV} above the valence band maximum), a high formation energy ($\sim$\SI{3.3}{eV} in the p-type regime, even under favorable B-rich conditions), and is highly compensated by the P$_\mathrm{B}$ donor. A similar situation exists for the dominant P$_\mathrm{B}$ donor under B-rich conditions. It is therefore unlikely that these native defects can be responsible for the carrier concentrations between \SI{e18}{cm^{-3}} and \SI{e21}{cm^{-3}} observed in most studies of crystalline BP.

Extrinsic defect calculations~\cite{Varley2017} indicate that Si, C, and Be should be more effective substitutional dopants compared to intrinsic defects in BP. All these impurities can dope BP either n-type or p-type depending on whether substitution occurs on the B site (n-type) or on the P site (p-type). Just like the intrinsic defects, B-poor conditions favor n-type conductivity and B-rich conditions favor p-type conductivity.~\cite{Varley2017}
Takigawa et al.~\cite{Takigawa1974} and Kumashiro~\cite{Kumashiro1990} had already hypothesized an "autodoping" mechanism in BP due to diffusion of Si from the substrate into BP. Substantial Si impurity levels (\SI{e20}{cm^{-3}} at \SI{1050}{\celsius} growth temperature) were detected in BP by secondary mass ion spectroscopy (SIMS).~\cite{Kumashiro1990} The Si content generally increased with increasing growth temperature and was correlated with the carrier concentration in BP.~\cite{Kumashiro1990} In general, the highest carrier concentrations reported in previous studies (above \SI{e19}{cm^{-3}}) have been achieved in the cases where BP was deposited on a Si-containing substrate and at high temperatures.~\cite{Chu1971,Iwami1975,Shohno1974} As another potential source of contamination, we note that the highest purity of elemental boron sources available from commercial suppliers is often only 99.9\%. Two different suppliers provided us with boron targets of this overall purity -- one with \SI{300}{ppm} Si, the other with \SI{10}{ppm} Si. The former target can potentially lead to carrier concentrations up to \SI{1.3 e19}{cm^{-3}} just due to its Si impurities.

The effect of possible C impurities on the electrical properties of BP has not been discussed in the experimental literature, although C from organic substances and cleaning agents can easily contaminate growth setups. C impurities in solid material sources (including our B sputter targets) are often not documented due to the very low C sensitivity factor of inductively coupled plasma mass spectrometry (ICP-MS), which is typically used for impurity analysis.~\cite{Houk2008} However, C is known to be a common impurity in elemental boron.~\cite{Greenwood1997,Yu2015a}

\begin{figure}[t!]
\centering%
\includegraphics[width=\columnwidth]{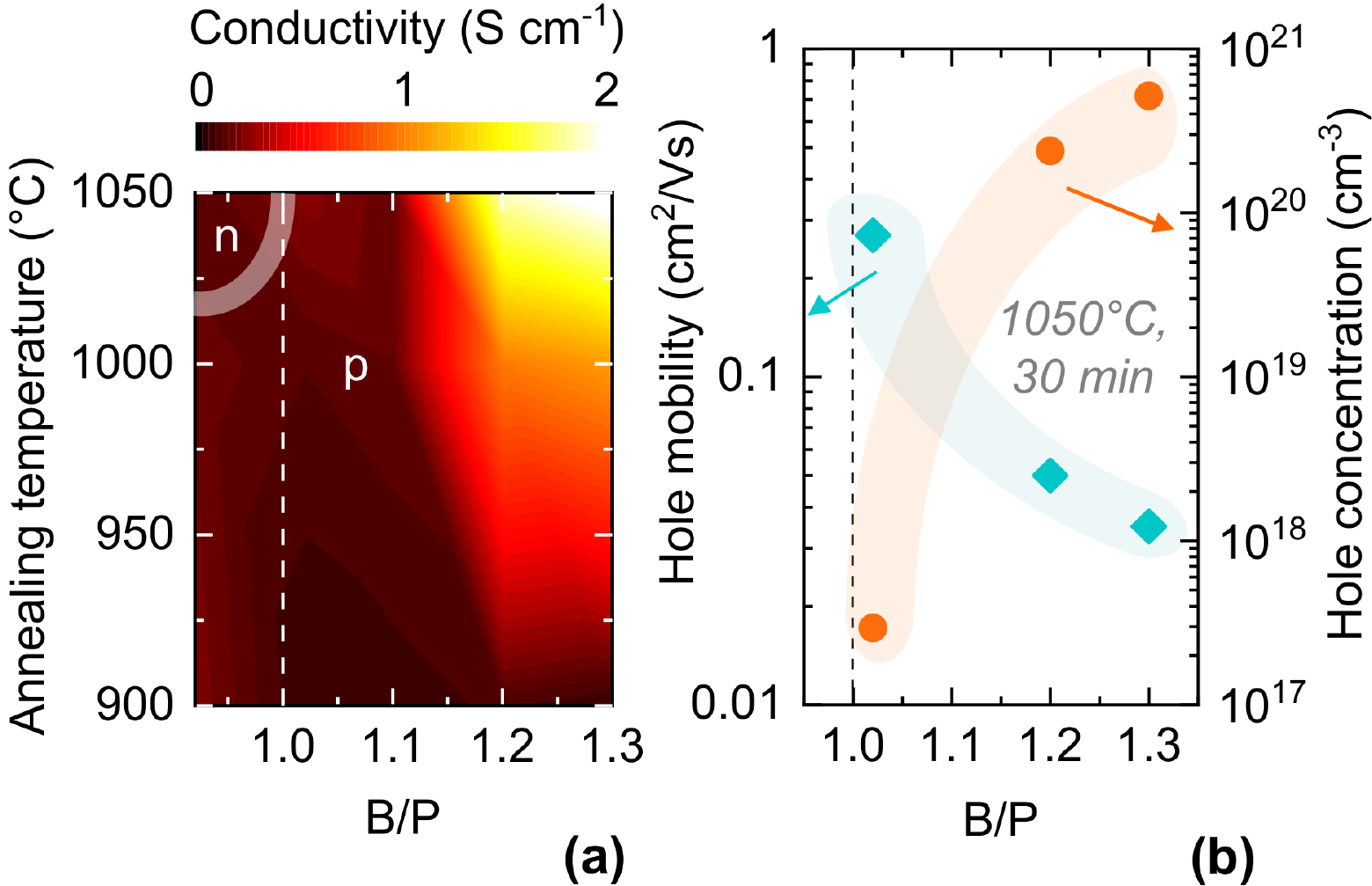}
\caption{Electrical properties of BP. (a) Conductivity of C-doped BP films as a function of annealing temperature and initial stoichiometry, with a constant annealing time of 30~min. Only the film with the lowest B/P ratio and annealed at the highest temperature is n-type. The other films are p-type according to Seebeck coefficient measurements. The data points are listed in Table~\ref{tab:SIconductivity_data}, Supporting Information. (b): Hole mobility and concentration for three films annealed at \SI{1050}{\celsius} as a function of initial composition. No reliable Hall effect measurements were possible for the other films.}
\label{fig:electrical}
\end{figure}

\subsubsection{C-doped BP with short post-annealing}

The electrical conductivity of BP films annealed for 30~min with 2.5\textendash 2.7~at.\% C dopant content is shown in Fig.~\ref{fig:electrical}(a) as a function of annealing temperature and initial B/P stoichiometry. N-type conductivity is only observed for B-poor films at high annealing temperatures, otherwise the films are p-type. The p-type conductivity increases for increasing B/P ratio, as expected from theory,~\cite{Varley2017} and also with increasing annealing temperature (Fig.~\ref{fig:electrical}(a)). The maximum p-type conductivity of \SI{2.1}{S/cm} is achieved for an annealing temperature of \SI{1050}{\celsius} and $\mathrm{B/P} = 1.3$. This conductivity value is in the typically reported range for BP single crystals and crystalline films.~\cite{Stone1960,Wang1964,Chu1971,Iwami1975,Kato1977,Kumashiro1990}

We obtained reliable Hall measurements only on three films annealed at \SI{1050}{\celsius} (Fig.~\ref{fig:electrical}(b)) using a Hall setup designed for low-mobility samples. These measurements already reveal important differences between our nanocrystalline, C-doped films and previously reported crystalline films.
The hole concentration in our B-rich films is well above \SI{e20}{cm^{-3}}, significantly higher than in any previous report. As expected, the hole concentration drops by several orders of magnitude when moving from B-rich composition to the nominal  $\mathrm{B/P} = 1$ stoichiometry (Fig.~\ref{fig:electrical}(b)). The hole mobility is \SI{0.27}{cm^2/Vs} near the stoichiometric B/P ratio and it drops by one order of magnitude when moving towards B-rich stoichiometries. An alternative non-contact technique (THz spectroscopy) indicates that the sum of electron and hole mobilities in the nearly stoichiometric BP film is about \SI{0.4}{cm^2/Vs} (Fig.~\ref{fig:SITHz}, Supporting Information). Due to the very small crystallite size in these films ($\sim$\SI{3}{nm}), we expect THz spectroscopy to be sensitive to the same scattering mechanisms as Hall effect measurements, hence the consistency between the two results. These mobility values are compatible with the only measurement on non-epitaxial, polycrystalline BP that we are aware of (\SI{0.13}{cm^2/Vs}).~\cite{Goossens1989} Hole mobilities above \SI{100}{cm^2/Vs} have, however, been demonstrated by several authors on single-crystalline BP.~\cite{Shohno1974,Wang1964} The low mobility of our stoichiometric film is likely due to the low crystalline quality obtained with short annealing times. The additional decrease in mobility for B-rich stoichiometries may be due to additional ionized impurity scattering from the C$_\mathrm{P}$ acceptors and/or to the increased concentration of other crystallographic defects caused by higher nonstoichiometry.

\begin{figure}[t!]
\centering%
\includegraphics[width=\columnwidth]{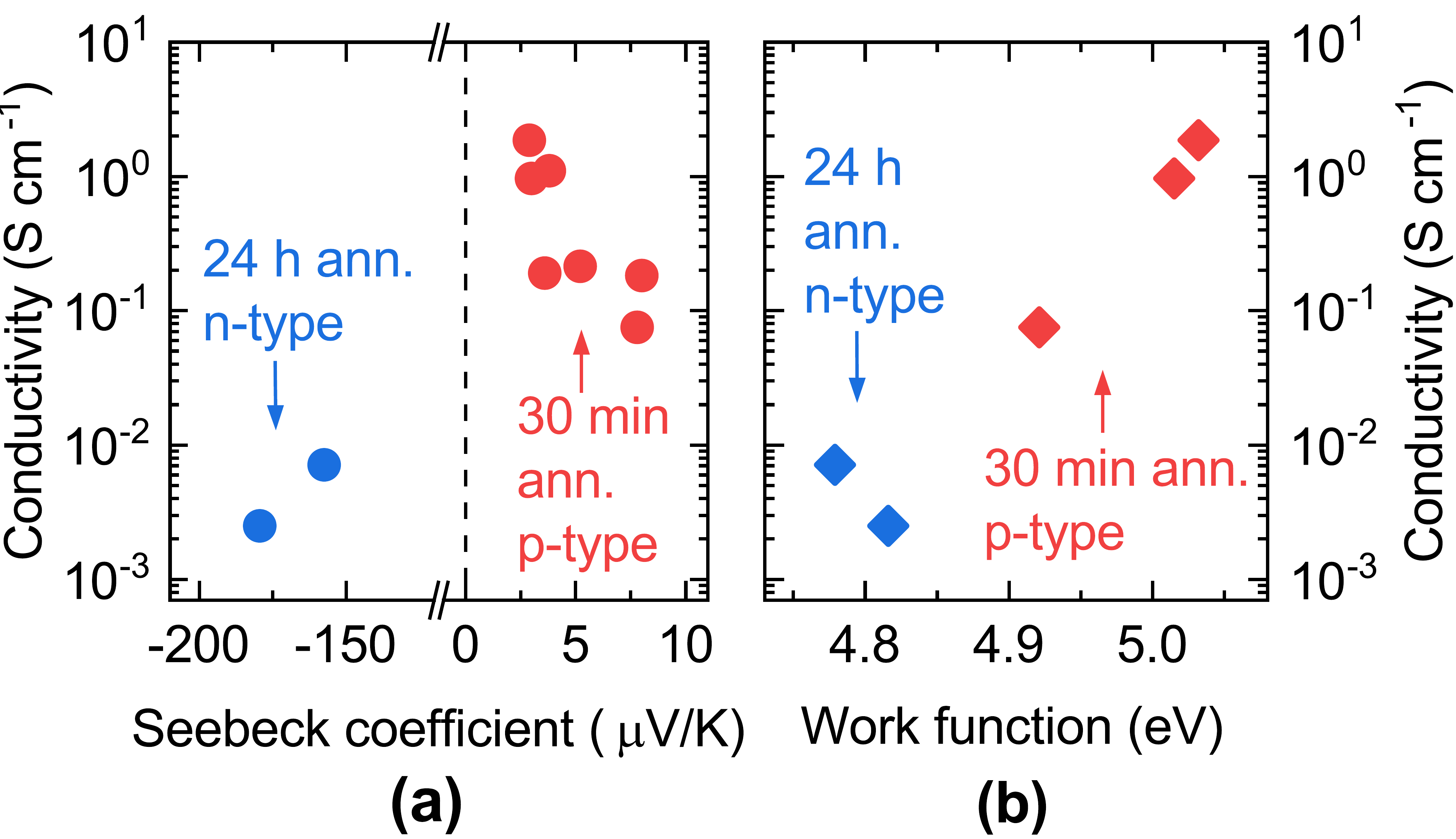}
\caption{Seebeck coefficient (a) and work function (b) of C-doped BP films with different doping types. Films annealed for 24~h generally have weak n-type doping (large negative Seebeck coefficients). Films annealed for 30~min are generally have higher p-type doping (small positive Seebeck coefficients). Despite type inversion, the surface Fermi level (work function) is pinned within a narrow region close to midgap.}
\label{fig:conductivity_seebeck_wf}
\end{figure}

\subsubsection{C-doped BP with long post-annealing} \label{sec:long}
The conductivities of films annealed for 24~h with sacrificial P powder and $\sim$2.5\% C doping are in the \SI{e-3}\textendash \SI{e-2}{S/cm} range regardless of initial stoichiometry. Their conductivity might be underestimated due to a high density of cracks (Fig.~\ref{fig:sem}(b)). Nevertheless, Seebeck coefficients in these films are generally much higher in magnitude than the Seebeck coefficients of the films annealed for a shorter time (Fig.~\ref{fig:conductivity_seebeck_wf}(a)). Due to the inverse relationship between Seebeck coefficient and carrier concentration, it is then likely that the films annealed for 24~h have much lower carrier concentrations. Another important difference is that films annealed for 24~h have negative Seebeck coefficients, indicating n-type conductivity, unlike the majority of the films annealed for 30~min (Fig.~\ref{fig:electrical}(a)). In the following paragraphs we present a hypothesis that could account for the observed differences.

We find that the depth-averaged composition of a BP film after a 30~min anneal at \SI{950}{\celsius} changes from $\mathrm{B/P} = 1.20$ to $\mathrm{B/P} = 0.99$. Thus, post-annealing in a P-containing atmosphere tends to compensate the original nonstoichiometry and produces nearly stoichiometric BP ($\mathrm{B/P} = 1$). This compensation process is apparently already completed after 30~min. Yet, the measured electrical properties still strongly depend on the B/P ratio of the original as-deposited films (Fig.~\ref{fig:electrical}). Hence, the high hole concentrations achieved by annealing B-rich BP for a short time can be considered a non-equilibrium property. In the initial stages of the annealing process, both C and P diffuse into B-rich amorphous BP precursors. Under B-rich conditions, formation of C$_\mathrm{P}$ acceptor defects is favorable, causing p-type conductivity. For extended annealing times, B-rich conditions no longer exist so the thermodynamically stable concentration of C$_\mathrm{P}$ acceptors is much lower than the non-equilibrium C$_\mathrm{P}$ concentration achieved in the initial annealing stages. Thus, a significant decrease in the C$_\mathrm{P}$ concentration is expected, but this decrease might take a long time to realize because (i) existing B-C bonds have to be broken, and (ii) BP becomes more and more crystalline with time, with a corresponding increase of kinetic barriers. As mentioned above, the final defect concentrations achieved after 24~h under our experimental conditions lead to weak n-type conductivity. The likely reason is that BP slowly becomes slightly P-rich in a P-containing atmosphere. This is compatible with the observation of elemental P secondary phases after long annealing times (Fig.~\ref{fig:sem}(b)).

\subsubsection{Work function}
According to Kelvin probe measurements, the work function of our BP films only spans a limited range (\SI{4.75}\textendash \SI{5.05}{eV}, Fig.~\ref{fig:conductivity_seebeck_wf}(b)), despite the fact that some films annealed for 30~min are degenerately p-type doped and the films annealed for 24~h are n-type. The ionization potential and electron affinity of CVD BP were previously determined as \SI{6.05}{eV} and \SI{4.05}{eV} respectively by electrochemical Mott-Schottky measurements.~\cite{Goossens1991} Thus, the surface Fermi level of our BP films is pinned around midgap despite their differences in doping type and concentration. Surface Fermi level pinning is a well-known phenomenon in air-exposed III-V semiconductor surfaces. The Fermi level of n-type and p-type GaAs exposed to oxygen is also pinned within a \SI{0.3}{eV} range around midgap.~\cite{Spicer1980} By analogy, we assume that a high density of surface states due to native oxide formation is responsible for the pinning effect in BP.

\subsubsection{C versus Si doping in sputtered BP}
\begin{figure}[t!]
\centering%
\includegraphics[width=\columnwidth]{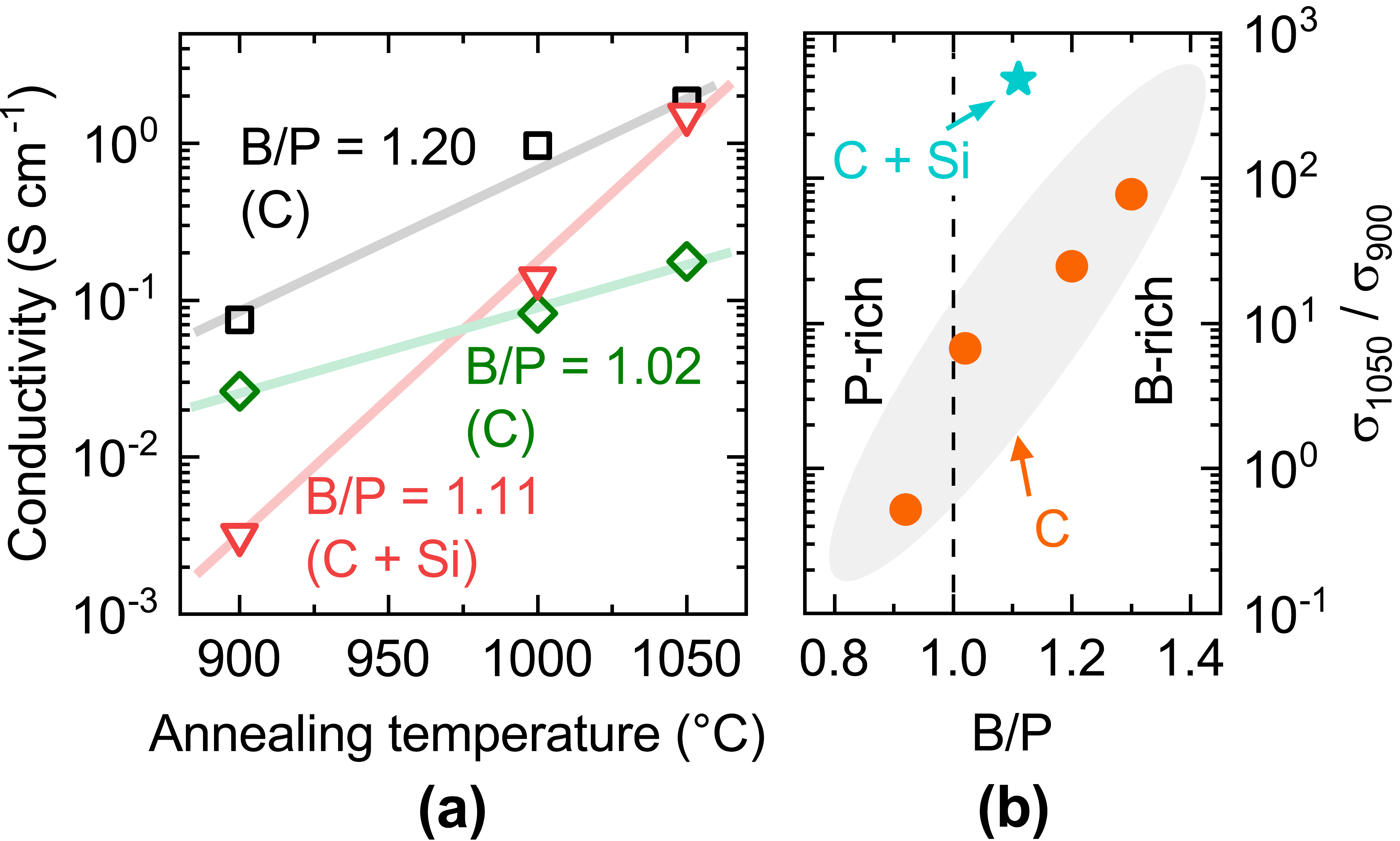}
\caption{Trends in the electrical properties of BP. (a) Conductivity of BP:Si,C films as a function of annealing temperature. The conductivity of two BP:C films is shown for comparison (same data as in Fig.~\ref{fig:electrical}(a)). In all film series, the conductivity has an approximately exponential dependence on annealing temperature. (b) Ratio between the conductivity at the highest annealing temperature ($\sigma_{1050}$) and at the lowest annealing temperature ($\sigma_{900}$). All data points used to plot these figures are listed in Table~\ref{tab:SIconductivity_data}, Supporting Information.}
\label{fig:silicon}
\end{figure}

To draw a preliminary comparison between C and Si as potential dopants in BP, we deposited a BP:Si film by reactive cosputtering of a B and a Si target. This film has $\mathrm{B/P} = 1.11$ and a Si content of approximately 2~at.\%. Different pieces of this film were annealed for 30~min and co-doped with $\sim$2.5~at.\%~C. We refer to these films as BP:Si,C. P-type conductivity is observed in all the BP:Si,C films after post-annealing. At \SI{1050}{\celsius} annealing temperature, the conductivity of BP:Si,C is about the same as the conductivity of a BP:C film with $\mathrm{B/P} = 1.20$ and much higher than the conductivity of a BP:C film with $\mathrm{B/P} = 1.02$ (Fig.~\ref{fig:silicon}(a)). The conductivity of all the films (BP:C and BP:Si,C) annealed for 30~min generally has an exponential dependence on annealing temperature (Fig.~\ref{fig:silicon}(a)). In Fig.~\ref{fig:silicon}(b), we plot the ratio $\sigma_{1050}/\sigma_{900}$ between the conductivity obtained after annealing at \SI{1050}{\celsius} and after annealing at\SI{900}{\celsius}, as a function of the B/P ratio. This quantity is a measure of the conductivity increase with increasing annealing temperature. For BP:C films, $\sigma_{1050}/\sigma_{900}$ increases with increasing B/P ratio (Fig.~\ref{fig:silicon}(b)). For the BP:Si,C series, $\sigma_{1050}/\sigma_{900}$ is much higher than expected for a BP:C film of the same B/P ratio (Fig.~\ref{fig:silicon}(b)).

A possible interpretation of the generally increasing conductivity with annealing temperature is dopant activation, i.e., the thermally-activated incorporation of C$_\mathrm{P}$ and Si$_\mathrm{P}$ acceptors in the BP lattice. To explain the higher $\sigma_{1050}/\sigma_{900}$ ratio in BP:Si,C films, one could invoke a higher activation energy for Si diffusion in BP or a higher defect formation energy for the Si$_\mathrm{P}$ defect.
However, these hypotheses do not explain why the BP:Si,C film has a much lower conductivity than all other BP:C films at the lowest annealing temperature (Table~\ref{tab:SIconductivity_data}, Supporting Information). It also does not explain why the $\sigma_{1050}/\sigma_{900}$ ratio for BP:C films increases significantly with increasing B/P ratio. It is important to reiterate that the B/P ratios quoted in this paper are measured before annealing. As discussed above, post-annealing in a P atmosphere tends to restore the stoichiometric B/P ratio of 1. At high annealing temperatures, we expect that B-rich conditions can be maintained for a longer time in films with high initial B/P ratios, explaining their higher $\sigma_{1050}/\sigma_{900}$ ratio.

The observation that the Si-codoped films are the least conductive at low annealing temperature could be rationalized by assuming that Si$_\mathrm{P}$ defects form more easily than C$_\mathrm{P}$ defects, but that they are less easily ionized by electrons in the valence band. With this hypothesis, Si$_\mathrm{P}$ defects would yield a lower concentration of free holes at room temperature compared to an equivalent concentration of C$_\mathrm{P}$ defects. In fact, this is exactly what was found by first-principles defect calculations.~\cite{Varley2017} The C$_\mathrm{P}$ acceptor has a higher formation enthalpy than Si$_\mathrm{P}$ at fixed chemical potential (\SI{3.0}{eV} versus \SI{1.7}{eV}). However, C$_\mathrm{P}$ has a much shallower level than Si$_\mathrm{P}$ (\SI{0.03}{eV} versus \SI{0.12}{eV} above the valence band maximum).

\section{Discussion}
It is relatively straightforward to deposit smooth, amorphous BP in a wide range of compositions by reactive sputtering. Nevertheless, this processing route has some disadvantages with respect to CVD. First, B is a very hard element with a much lower sputter yield than the soft metals used in n-type TCMs (e.g., Zn, In, Sn). Since high sputter powers are not viable due to the low thermal conductivity of B, low deposition rates in reactive sputtering of B seem unavoidable. A BP compound target is also expected to have a low sputter yield but it may allow for higher powers due its high thermal conductivity.~\cite{Kang2017a} A potential advantage of reactively sputtered BP is that the as-deposited amorphous films have a relatively low surface roughness of around \SI{2}{nm} RMS for \SI{100}\textendash \SI{200}{nm}-thick films (Table~\ref{tab:SIellipsometry_data}, Supporting Information). BP has been proposed as a material for extreme UV optics but the large roughness of CVD films was found to be a limiting factor.~\cite{Huber2016} Thus, reactive sputtering may be a more promising route for this particular application. 

Nanocrystalline BP films with small ($\sim$\SI{3}{nm}) crystallite size are not sufficiently transparent for TCM applications due to strong absorption above its nominally indirect band gap. High crystalline quality is thus an important requirement for BP and for other candidate TCMs with indirect gaps in the visible. This requirement is much less stringent in direct-gap n-type oxide TCMs, which often have very small grain sizes.~\cite{Crovetto2016a,Morales-Masis2017} Unfortunately, reactively sputtered BP films on fused silica are still amorphous even at a substrate temperature of \SI{800}{\celsius}. The crystallinity of BP films can be improved by post-annealing in a P-containing atmosphere but the crystallite size seems to be limited by the particle size of the as-deposited films, without significant coalescence between these particles. The limiting crystallite size obtained by long anneals at high temperatures is in the \SI{30}\textendash \SI{40}{nm} range. This crystallite size might be sufficient to prevent strong indirect absorption, but we could not draw any conclusion in this regard because our crystalline films have spurious phases at the surface (Fig.~\ref{fig:sem}(b)). Significant light scattering from these phases, as well as their unknown optical properties, makes it difficult to determine the intrinsic absorption coefficient of the underlying BP film.

With short annealing times, very high hole concentrations can be achieved in nanocrystalline BP under B-rich conditions using extrinsic dopants. Higher hole concentrations are realized with C dopants than with Si dopants, probably because C$_\mathrm{P}$ is significantly shallower than Si$_\mathrm{P}$. Since the conductivity of BP films without extrinsic doping was below \SI{e-3}{S/cm} under all the conditions investigated in this study, we believe that some of the previously reported high carrier concentrations in BP may be due to Si and/or C incorporation from the substrate or other sources.

Long annealing times have the advantage of better BP crystalline quality and the potential for higher optical transmission. However, we encountered three issues. First, a higher density of cracks (Fig.~\ref{fig:sem}(b)). Second, formation of secondary phases which we identified as elemental phosphorus (Fig.~\ref{fig:SIauger_map}, Supporting Information). Third, the tendency of B-rich films to reach their stoichiometric B/P ratio by incorporating P from the evaporated P powder (Sec.~\ref{sec:long}). This tendency limits the achievable hole concentration, because acceptor defects have higher formation energies under stoichiometric conditions. The issue with cracks might be addressed by a different substrate with a higher thermal expansion coefficient. The two other issues could be addressed by optimizing the amount of sacrificial P powder and employing a thermal gradient to avoid condensation of excess P on the BP films. An annealing system with a continuous PH$_3$ flow might be a more practical solution to control P incorporation in the film.

\section{Conclusion and outlook}
Amorphous BP films can be deposited by reactive sputtering in a wide range of compositions. However, they require a post-annealing step at \SI{1000}{\celsius} or above in a P-containing atmosphere for crystallization and dopant activation. Our study points to C as a more effective p-type dopant than either Si or native defects in BP. Incorporation of 2.5\% C in BP led to the highest hole concentration reported to date for p-type BP (\SI{5e20}{cm^{-3}}). While an initial B-rich composition generally favors p-type conductivity, it was also possible to obtain n-type conductivity in B-poor films. Thus, we confirm that BP is amenable to bipolar doping, a rare feature in wide band gap materials.

The hole mobilities and optical transmission of our films were too low for TCM applications. The most likely cause is the relatively poor crystalline quality of BP films annealed for a relatively short time. In general, incomplete crystallization will penalize any candidate transparent conductor with an indirect gap in the visible. The reason is that participation of phonons in indirect transitions is not required in the absence of long-range order, so the absorption coefficient will increase with respect to the case of a perfect crystal.
It may be possible to improve the optoelectronic properties through longer anneals with optimized process parameters. Nevertheless, single-step deposition processes at higher P partial pressures are more likely to achieve higher crystalline quality and seem to be a more practical solution in general. Based on previously reported mobilities and absorption coefficient in crystalline BP, as well as the high hole concentrations demonstrated in this study, we argue that BP:C has the potential to achieve high TCM figures of merit and certainly deserves further investigation. Depositing C-doped BP by CVD using e.g., a CH$_4$ doping source, might be a promising strategy. In particular, it will be important to clarify if nonepitaxial, polycrystalline BP can achieve similar hole mobilities and optical transparency as single-crystalline films.

High process temperatures seem unavoidable for obtaining high crystalline quality and dopant incorporation in BP, so it is likely that BP TCMs can only find applications as the first layer in a device stack. Still, a TCM on a transparent substrate is the initial layer in many important optoelectronic devices, such as CdTe and perovskite solar cells,~\cite{Woods-Robinson2020,Saliba2018} so device applications are in principle possible. Since post-annealed BP films on fused silica substrates are prone to cracking, alternative refractory and transparent substrates with a higher thermal expansion coefficient might be necessary. BP is itself a refractory material, and this property could be particularly useful as a transparent contact in thin-film optoelectronic technologies requiring high-temperature processing in other functional layers, such as the emerging class of chalcogenide perovskites.~\cite{Crovetto2019} As an additional advantage, the thermal conductivity of BP is among the highest known for any material.~\cite{Kang2017a} In bifacial solar cells lacking a continuous metallic contact and only consisting of materials with low thermal conductivity, BP contacts could have an additional thermal management function. Finally, the possibility of bipolar doping is encouraging for possible applications in transparent electronics.

\section{Experimental details}

\subsection{Film deposition}
Amorphous BP films were deposited on fused silica and crystalline Si substrates by reactive radio-frequency (RF) magnetron sputtering of B targets in an Ar/PH$_3$ atmosphere. We obtained the desired Ar/PH$_3$ ratio by tuning the relative flow rate of a diluted PH$_3$ source (5\% in Ar) and a pure Ar source. Due to the very low sputter yield of B, the films were grown under simultaneous sputtering of two 2" B targets from separate magnetron sources. We used a bare 0.25"-thick B target (99.9\% purity, Kurt J. Lesker) and a 0.125"-thick B target In-bonded to a Cu backing plate (99.9\% purity, Advanced Engineering Materials). The two targets were confocally oriented with respect to the middle of the substrate. The gas inlet inside the deposition chamber was at approximately the same distance from the two targets and from the substrate. The substrates were clamped to a metallic platen, which was heated by infrared lamps. The quoted temperatures are those measured on the metallic platen after the 30~min stabilization time that preceded each deposition. The BP films employed for post-annealing experiments were deposited at room temperature with a total pressure of \SI{5}{mTorr}, a PH$_3$ partial pressure in the \SI{0.02}\textendash \SI{0.06}{mTorr} range, an RF power of \SI{40}{W} for each target, and a target-substrate distance of \SI{10}{cm}. A description of safety measures adopted by the process operators is available elsewhere.~\cite{Schnepf2021}

\subsection{Post-annealing}
Ampoules for annealing treatments were made from 1220~mm long, 12~mm outer diameter, and 2~mm wall thickness fused silica tubing (Technical Glass Products, OH, USA). The tubes were cut into 400~mm sections using a diamond saw. The open ended 400~mm tube sections were rinsed using water followed by isopropyl alcohol to remove any residual water. Tube sections were then dried using dry compressed air. An oxy-propylene torch was used to pull the 400~mm tube sections into half-ampoules (200~mm sections with a single closed end). As-deposited BP films were cut into $7 \times 12$~mm$^2$ pieces for post-annealing. Two such samples were placed in each half-ampoule. Care was taken to ensure that there was no overlap between the two samples inside of the ampoules. Unless otherwise specified, samples were annealed in a P-containing atmosphere by loading sacrificial red phosphorus powder (3~mg) in the ampoule. Ampoules were then sealed in vacuum after three evacuation ($<$100~mTorr) and purging (Ar) cycles, with a final sealed ampoule length of 120~mm. Under these conditions, the total phosphorus pressure at \SI{1000}{\celsius} is estimated as 500~Torr and is expected to consist of gaseous P$_4$ and P$_2$ species in approximately equal concentrations.~\cite{Bock1984} For the annealing experiments without P powder, the ampoules were sealed under 400~Torr of Ar after the last evacuation step. The ampoules were placed horizontally in a tube furnace (Thermcraft Inc., NC, USA) as close to the control thermocouple as possible. A fixed heating and cooling ramp rate of 20$^\circ$C/min was used while the soak time and temperatures were changed as a variable.

\subsection{Characterization}
X-ray diffraction measurements were conducted with a Bruker D8 diffractometer using Cu K$_\alpha$ radiation and a 2D detector. To cover the desired $2\theta$ range, three frames were collected with the incidence angle $\omega$ fixed at 10$^\circ$, 22.5$^\circ$, and 35$^\circ$ and the detector center fixed at $2\theta$ values of 35$^\circ$, 60$^\circ$, and 85$^\circ$ respectively. The diffraction intensity at each $2\theta$ angle was integrated over the $\chi$ range measured by the 2D detector. Structural analysis by XRD was complemented by Raman spectroscopy (Renishaw inVia) at \SI{532}{nm} excitation and \SI{40}{W/mm^2} power density.

Film composition was determined by Auger electron spectroscopy (AES, Physical Electronics 710 scanning Auger nanoprobe) with a \SI{10}{kV}, \SI{10}{nA} primary beam and after removing the native oxide and adsorbed species with an Ar$^+$ beam (\SI{2}{kV}). The default sensitivity factors for numerically differentiated B-\textit{KLL} and P-\textit{LMM} peaks were used for quantification, which was performed using Physical Electronics MultiPak v9.6.1.7. Unless other-wise specified, the B/P ratios quoted in the article refer to atomic composition of the films before annealing. Scanning electron microscopy (SEM) images were taken with a Hitachi S-3400N instrument with a field emission gun and \SI{5}{kV} beam voltage.

Film thickness and optical properties were determined by variable-angle spectroscopic ellipsometry measurements using a J.A. Woollam Co. M-2000 rotating-compensator ellipsometer. Ellipsometry spectra were fitted with a substrate/film/roughness layer model in the CompleteEase software (J.A. Woollam). The fitting parameters were the thicknesses of the film and roughness layers, as well as the dielectric function of the film layer. The latter was represented by a Kramers-Kronig-consistent b-spline function with 0.3 nodes/eV. Optical transmission in the visible and near IR was measured in a home-built setup.

Electrical conductivity was measured on the plane of the substrate by a collinear four-point-probe. Hall carrier concentration and mobility were measured with a Hall measurement station for low-mobility samples (FastHall, LakeShore Cryotronics) in the van der Pauw configuration. The results were validated by reversing the magnetic field and changing the measurement current. In case of inconsistencies, the results were discarded. The ellipsometry-determined film thickness was used in the derivation of the electrical properties.

The Seebeck coefficient was measured in a home-built setup using In contacts and four temperature differences in the vicinity of room temperature. After checking that the thermovoltage depended linearly on the temperature difference, the Seebeck coefficient of BP was determined as the slope of the linear curve minus the Seebeck coefficient of the In contacts. The work function was measured with a SKP SPV LE 450 Kelvin probe (KP Technology) calibrated with a standard Au sample.

\section*{Acknowledgements}
This project has received funding from the European Union’s Horizon 2020 research and innovation programme under the Marie Sk\l odowska-Curie grant agreement No 840751 (synthesis, characterization, and analysis work).
This work was authored in part at the National Renewable Energy Laboratory, operated by Alliance for Sustainable Energy, LLC, for the U.S. Department of Energy (DOE) under Contract No. DE-AC36-08GO28308. Funding supporting development of synthesis and characterization equipment (A.C.T., R.R.S., A.Z.) was provided by the Office of Science, Office of Basic Energy Sciences. J.M.A. and E.S.T. acknowledge NSF DMR award 1555340. A.C. thanks Joel Basile Varley and Geoffroy Hautier for useful discussions.

\bibliography{library}

\begin{thebibliography}{70}%
\makeatletter
\providecommand \@ifxundefined [1]{%
 \@ifx{#1\undefined}
}%
\providecommand \@ifnum [1]{%
 \ifnum #1\expandafter \@firstoftwo
 \else \expandafter \@secondoftwo
 \fi
}%
\providecommand \@ifx [1]{%
 \ifx #1\expandafter \@firstoftwo
 \else \expandafter \@secondoftwo
 \fi
}%
\providecommand \natexlab [1]{#1}%
\providecommand \enquote  [1]{``#1''}%
\providecommand \bibnamefont  [1]{#1}%
\providecommand \bibfnamefont [1]{#1}%
\providecommand \citenamefont [1]{#1}%
\providecommand \href@noop [0]{\@secondoftwo}%
\providecommand \href [0]{\begingroup \@sanitize@url \@href}%
\providecommand \@href[1]{\@@startlink{#1}\@@href}%
\providecommand \@@href[1]{\endgroup#1\@@endlink}%
\providecommand \@sanitize@url [0]{\catcode `\\12\catcode `\$12\catcode
  `\&12\catcode `\#12\catcode `\^12\catcode `\_12\catcode `\%12\relax}%
\providecommand \@@startlink[1]{}%
\providecommand \@@endlink[0]{}%
\providecommand \url  [0]{\begingroup\@sanitize@url \@url }%
\providecommand \@url [1]{\endgroup\@href {#1}{\urlprefix }}%
\providecommand \urlprefix  [0]{URL }%
\providecommand \Eprint [0]{\href }%
\providecommand \doibase [0]{http://dx.doi.org/}%
\providecommand \selectlanguage [0]{\@gobble}%
\providecommand \bibinfo  [0]{\@secondoftwo}%
\providecommand \bibfield  [0]{\@secondoftwo}%
\providecommand \translation [1]{[#1]}%
\providecommand \BibitemOpen [0]{}%
\providecommand \bibitemStop [0]{}%
\providecommand \bibitemNoStop [0]{.\EOS\space}%
\providecommand \EOS [0]{\spacefactor3000\relax}%
\providecommand \BibitemShut  [1]{\csname bibitem#1\endcsname}%
\let\auto@bib@innerbib\@empty
\bibitem [{\citenamefont {Kang}\ \emph {et~al.}(2017)\citenamefont {Kang},
  \citenamefont {Wu},\ and\ \citenamefont {Hu}}]{Kang2017a}%
  \BibitemOpen
  \bibfield  {author} {\bibinfo {author} {\bibfnamefont {J.~S.}\ \bibnamefont
  {Kang}}, \bibinfo {author} {\bibfnamefont {H.}~\bibnamefont {Wu}}, \ and\
  \bibinfo {author} {\bibfnamefont {Y.}~\bibnamefont {Hu}},\ }\href {\doibase
  10.1021/acs.nanolett.7b03437} {\bibfield  {journal} {\bibinfo  {journal}
  {Nano Letters}\ }\textbf {\bibinfo {volume} {17}},\ \bibinfo {pages} {7507}
  (\bibinfo {year} {2017})}\BibitemShut {NoStop}%
\bibitem [{\citenamefont {Stone}\ and\ \citenamefont {Hill}(1960)}]{Stone1960}%
  \BibitemOpen
  \bibfield  {author} {\bibinfo {author} {\bibfnamefont {B.}~\bibnamefont
  {Stone}}\ and\ \bibinfo {author} {\bibfnamefont {D.}~\bibnamefont {Hill}},\
  }\href {\doibase 10.1103/PhysRevLett.4.282} {\bibfield  {journal} {\bibinfo
  {journal} {Physical Review Letters}\ }\textbf {\bibinfo {volume} {4}},\
  \bibinfo {pages} {282} (\bibinfo {year} {1960})}\BibitemShut {NoStop}%
\bibitem [{\citenamefont {Shohno}\ \emph {et~al.}(1974)\citenamefont {Shohno},
  \citenamefont {Takigawa},\ and\ \citenamefont {Nakada}}]{Shohno1974}%
  \BibitemOpen
  \bibfield  {author} {\bibinfo {author} {\bibfnamefont {K.}~\bibnamefont
  {Shohno}}, \bibinfo {author} {\bibfnamefont {M.}~\bibnamefont {Takigawa}}, \
  and\ \bibinfo {author} {\bibfnamefont {T.}~\bibnamefont {Nakada}},\ }\href
  {\doibase 10.1016/0022-0248(74)90303-0} {\bibfield  {journal} {\bibinfo
  {journal} {Journal of Crystal Growth}\ }\textbf {\bibinfo {volume} {24-25}},\
  \bibinfo {pages} {193} (\bibinfo {year} {1974})}\BibitemShut {NoStop}%
\bibitem [{\citenamefont {Ginley}\ \emph {et~al.}(1983)\citenamefont {Ginley},
  \citenamefont {Baughman},\ and\ \citenamefont {Butler}}]{Ginley1983}%
  \BibitemOpen
  \bibfield  {author} {\bibinfo {author} {\bibfnamefont {D.~S.}\ \bibnamefont
  {Ginley}}, \bibinfo {author} {\bibfnamefont {R.~J.}\ \bibnamefont
  {Baughman}}, \ and\ \bibinfo {author} {\bibfnamefont {M.~A.}\ \bibnamefont
  {Butler}},\ }\href {\doibase 10.1149/1.2119509} {\bibfield  {journal}
  {\bibinfo  {journal} {Journal of The Electrochemical Society}\ }\textbf
  {\bibinfo {volume} {130}},\ \bibinfo {pages} {1999} (\bibinfo {year}
  {1983})}\BibitemShut {NoStop}%
\bibitem [{\citenamefont {Kumashiro}(1990)}]{Kumashiro1990}%
  \BibitemOpen
  \bibfield  {author} {\bibinfo {author} {\bibfnamefont {Y.}~\bibnamefont
  {Kumashiro}},\ }\href {\doibase 10.1557/JMR.1990.2933} {\bibfield  {journal}
  {\bibinfo  {journal} {Journal of Materials Research}\ }\textbf {\bibinfo
  {volume} {5}},\ \bibinfo {pages} {2933} (\bibinfo {year} {1990})}\BibitemShut
  {NoStop}%
\bibitem [{\citenamefont {Shi}\ \emph {et~al.}(2016)\citenamefont {Shi},
  \citenamefont {Li}, \citenamefont {Zhou}, \citenamefont {Wang}, \citenamefont
  {Chang}, \citenamefont {Zhang}, \citenamefont {Kako}, \citenamefont {Liu},\
  and\ \citenamefont {Ye}}]{Shi2016}%
  \BibitemOpen
  \bibfield  {author} {\bibinfo {author} {\bibfnamefont {L.}~\bibnamefont
  {Shi}}, \bibinfo {author} {\bibfnamefont {P.}~\bibnamefont {Li}}, \bibinfo
  {author} {\bibfnamefont {W.}~\bibnamefont {Zhou}}, \bibinfo {author}
  {\bibfnamefont {T.}~\bibnamefont {Wang}}, \bibinfo {author} {\bibfnamefont
  {K.}~\bibnamefont {Chang}}, \bibinfo {author} {\bibfnamefont
  {H.}~\bibnamefont {Zhang}}, \bibinfo {author} {\bibfnamefont
  {T.}~\bibnamefont {Kako}}, \bibinfo {author} {\bibfnamefont {G.}~\bibnamefont
  {Liu}}, \ and\ \bibinfo {author} {\bibfnamefont {J.}~\bibnamefont {Ye}},\
  }\href {\doibase 10.1016/j.nanoen.2016.08.041} {\bibfield  {journal}
  {\bibinfo  {journal} {Nano Energy}\ }\textbf {\bibinfo {volume} {28}},\
  \bibinfo {pages} {158} (\bibinfo {year} {2016})}\BibitemShut {NoStop}%
\bibitem [{\citenamefont {Mou}\ \emph {et~al.}(2019)\citenamefont {Mou},
  \citenamefont {Wu}, \citenamefont {Xie}, \citenamefont {Zhang}, \citenamefont
  {Ji}, \citenamefont {Huang}, \citenamefont {Wang}, \citenamefont {Luo},
  \citenamefont {Xiong}, \citenamefont {Tang},\ and\ \citenamefont
  {Sun}}]{Mou2019}%
  \BibitemOpen
  \bibfield  {author} {\bibinfo {author} {\bibfnamefont {S.}~\bibnamefont
  {Mou}}, \bibinfo {author} {\bibfnamefont {T.}~\bibnamefont {Wu}}, \bibinfo
  {author} {\bibfnamefont {J.}~\bibnamefont {Xie}}, \bibinfo {author}
  {\bibfnamefont {Y.}~\bibnamefont {Zhang}}, \bibinfo {author} {\bibfnamefont
  {L.}~\bibnamefont {Ji}}, \bibinfo {author} {\bibfnamefont {H.}~\bibnamefont
  {Huang}}, \bibinfo {author} {\bibfnamefont {T.}~\bibnamefont {Wang}},
  \bibinfo {author} {\bibfnamefont {Y.}~\bibnamefont {Luo}}, \bibinfo {author}
  {\bibfnamefont {X.}~\bibnamefont {Xiong}}, \bibinfo {author} {\bibfnamefont
  {B.}~\bibnamefont {Tang}}, \ and\ \bibinfo {author} {\bibfnamefont
  {X.}~\bibnamefont {Sun}},\ }\href {\doibase 10.1002/adma.201903499}
  {\bibfield  {journal} {\bibinfo  {journal} {Advanced Materials}\ }\textbf
  {\bibinfo {volume} {31}},\ \bibinfo {pages} {1903499} (\bibinfo {year}
  {2019})}\BibitemShut {NoStop}%
\bibitem [{\citenamefont {Huber}\ \emph {et~al.}(2016)\citenamefont {Huber},
  \citenamefont {Medvedev}, \citenamefont {Meyer-Ilse}, \citenamefont
  {Gullikson}, \citenamefont {Padavala}, \citenamefont {Edgar}, \citenamefont
  {Sturm}, \citenamefont {van~de Kruijs}, \citenamefont {Prendergast},\ and\
  \citenamefont {Bijkerk}}]{Huber2016}%
  \BibitemOpen
  \bibfield  {author} {\bibinfo {author} {\bibfnamefont {S.~P.}\ \bibnamefont
  {Huber}}, \bibinfo {author} {\bibfnamefont {V.~V.}\ \bibnamefont {Medvedev}},
  \bibinfo {author} {\bibfnamefont {J.}~\bibnamefont {Meyer-Ilse}}, \bibinfo
  {author} {\bibfnamefont {E.}~\bibnamefont {Gullikson}}, \bibinfo {author}
  {\bibfnamefont {B.}~\bibnamefont {Padavala}}, \bibinfo {author}
  {\bibfnamefont {J.~H.}\ \bibnamefont {Edgar}}, \bibinfo {author}
  {\bibfnamefont {J.~M.}\ \bibnamefont {Sturm}}, \bibinfo {author}
  {\bibfnamefont {R.~W.~E.}\ \bibnamefont {van~de Kruijs}}, \bibinfo {author}
  {\bibfnamefont {D.}~\bibnamefont {Prendergast}}, \ and\ \bibinfo {author}
  {\bibfnamefont {F.}~\bibnamefont {Bijkerk}},\ }\href {\doibase
  10.1364/OME.6.003946} {\bibfield  {journal} {\bibinfo  {journal} {Optical
  Materials Express}\ }\textbf {\bibinfo {volume} {6}},\ \bibinfo {pages}
  {3946} (\bibinfo {year} {2016})}\BibitemShut {NoStop}%
\bibitem [{\citenamefont {Varley}\ \emph {et~al.}(2017)\citenamefont {Varley},
  \citenamefont {Miglio}, \citenamefont {Ha}, \citenamefont {van Setten},
  \citenamefont {Rignanese},\ and\ \citenamefont {Hautier}}]{Varley2017}%
  \BibitemOpen
  \bibfield  {author} {\bibinfo {author} {\bibfnamefont {J.~B.}\ \bibnamefont
  {Varley}}, \bibinfo {author} {\bibfnamefont {A.}~\bibnamefont {Miglio}},
  \bibinfo {author} {\bibfnamefont {V.-A.}\ \bibnamefont {Ha}}, \bibinfo
  {author} {\bibfnamefont {M.~J.}\ \bibnamefont {van Setten}}, \bibinfo
  {author} {\bibfnamefont {G.-M.}\ \bibnamefont {Rignanese}}, \ and\ \bibinfo
  {author} {\bibfnamefont {G.}~\bibnamefont {Hautier}},\ }\href {\doibase
  10.1021/acs.chemmater.6b04663} {\bibfield  {journal} {\bibinfo  {journal}
  {Chemistry of Materials}\ }\textbf {\bibinfo {volume} {29}},\ \bibinfo
  {pages} {2568} (\bibinfo {year} {2017})}\BibitemShut {NoStop}%
\bibitem [{\citenamefont {Fioretti}\ and\ \citenamefont
  {Morales-Masis}(2020)}]{Fioretti2020}%
  \BibitemOpen
  \bibfield  {author} {\bibinfo {author} {\bibfnamefont {A.~N.}\ \bibnamefont
  {Fioretti}}\ and\ \bibinfo {author} {\bibfnamefont {M.}~\bibnamefont
  {Morales-Masis}},\ }\href {\doibase 10.1117/1.JPE.10.042002} {\bibfield
  {journal} {\bibinfo  {journal} {Journal of Photonics for Energy}\ }\textbf
  {\bibinfo {volume} {10}},\ \bibinfo {pages} {042002} (\bibinfo {year}
  {2020})}\BibitemShut {NoStop}%
\bibitem [{\citenamefont {Willis}\ and\ \citenamefont
  {Scanlon}(2021)}]{Willis2021}%
  \BibitemOpen
  \bibfield  {author} {\bibinfo {author} {\bibfnamefont {J.}~\bibnamefont
  {Willis}}\ and\ \bibinfo {author} {\bibfnamefont {D.~O.}\ \bibnamefont
  {Scanlon}},\ }\href {\doibase 10.1039/D1TC02547C} {\bibfield  {journal}
  {\bibinfo  {journal} {Journal of Materials Chemistry C}\ } (\bibinfo {year}
  {2021}),\ 10.1039/D1TC02547C}\BibitemShut {NoStop}%
\bibitem [{\citenamefont {Takigawa}\ \emph {et~al.}(1974)\citenamefont
  {Takigawa}, \citenamefont {Hirayama},\ and\ \citenamefont
  {Shohno}}]{Takigawa1974}%
  \BibitemOpen
  \bibfield  {author} {\bibinfo {author} {\bibfnamefont {M.}~\bibnamefont
  {Takigawa}}, \bibinfo {author} {\bibfnamefont {M.}~\bibnamefont {Hirayama}},
  \ and\ \bibinfo {author} {\bibfnamefont {K.}~\bibnamefont {Shohno}},\ }\href
  {\doibase 10.1143/JJAP.13.411} {\bibfield  {journal} {\bibinfo  {journal}
  {Japanese Journal of Applied Physics}\ }\textbf {\bibinfo {volume} {13}},\
  \bibinfo {pages} {411} (\bibinfo {year} {1974})}\BibitemShut {NoStop}%
\bibitem [{\citenamefont {Iwami}\ \emph {et~al.}(1975)\citenamefont {Iwami},
  \citenamefont {Tohda},\ and\ \citenamefont {Kawabe}}]{Iwami1975}%
  \BibitemOpen
  \bibfield  {author} {\bibinfo {author} {\bibfnamefont {M.}~\bibnamefont
  {Iwami}}, \bibinfo {author} {\bibfnamefont {T.}~\bibnamefont {Tohda}}, \ and\
  \bibinfo {author} {\bibfnamefont {K.}~\bibnamefont {Kawabe}},\ }\href
  {\doibase 10.1002/eej.4390950304} {\bibfield  {journal} {\bibinfo  {journal}
  {Electrical Engineering in Japan}\ }\textbf {\bibinfo {volume} {95}},\
  \bibinfo {pages} {19} (\bibinfo {year} {1975})}\BibitemShut {NoStop}%
\bibitem [{\citenamefont {Archer}\ \emph {et~al.}(1964)\citenamefont {Archer},
  \citenamefont {Koyama}, \citenamefont {Loebner},\ and\ \citenamefont
  {Lucas}}]{Archer1964}%
  \BibitemOpen
  \bibfield  {author} {\bibinfo {author} {\bibfnamefont {R.~J.}\ \bibnamefont
  {Archer}}, \bibinfo {author} {\bibfnamefont {R.~Y.}\ \bibnamefont {Koyama}},
  \bibinfo {author} {\bibfnamefont {E.~E.}\ \bibnamefont {Loebner}}, \ and\
  \bibinfo {author} {\bibfnamefont {R.~C.}\ \bibnamefont {Lucas}},\ }\href
  {\doibase 10.1103/PhysRevLett.12.538} {\bibfield  {journal} {\bibinfo
  {journal} {Physical Review Letters}\ }\textbf {\bibinfo {volume} {12}},\
  \bibinfo {pages} {538} (\bibinfo {year} {1964})}\BibitemShut {NoStop}%
\bibitem [{\citenamefont {Ha}\ \emph {et~al.}(2020)\citenamefont {Ha},
  \citenamefont {Karasulu}, \citenamefont {Maezono}, \citenamefont {Brunin},
  \citenamefont {Varley}, \citenamefont {Rignanese}, \citenamefont
  {Monserrat},\ and\ \citenamefont {Hautier}}]{Ha2020}%
  \BibitemOpen
  \bibfield  {author} {\bibinfo {author} {\bibfnamefont {V.-A.}\ \bibnamefont
  {Ha}}, \bibinfo {author} {\bibfnamefont {B.}~\bibnamefont {Karasulu}},
  \bibinfo {author} {\bibfnamefont {R.}~\bibnamefont {Maezono}}, \bibinfo
  {author} {\bibfnamefont {G.}~\bibnamefont {Brunin}}, \bibinfo {author}
  {\bibfnamefont {J.~B.}\ \bibnamefont {Varley}}, \bibinfo {author}
  {\bibfnamefont {G.-M.}\ \bibnamefont {Rignanese}}, \bibinfo {author}
  {\bibfnamefont {B.}~\bibnamefont {Monserrat}}, \ and\ \bibinfo {author}
  {\bibfnamefont {G.}~\bibnamefont {Hautier}},\ }\href {\doibase
  10.1103/PhysRevMaterials.4.065401} {\bibfield  {journal} {\bibinfo  {journal}
  {Physical Review Materials}\ }\textbf {\bibinfo {volume} {4}},\ \bibinfo
  {pages} {065401} (\bibinfo {year} {2020})},\ \Eprint
  {http://arxiv.org/abs/2004.05390} {arXiv:2004.05390} \BibitemShut {NoStop}%
\bibitem [{\citenamefont {Schroten}\ \emph {et~al.}(1998)\citenamefont
  {Schroten}, \citenamefont {Goossens},\ and\ \citenamefont
  {Schoonman}}]{Schroten1998}%
  \BibitemOpen
  \bibfield  {author} {\bibinfo {author} {\bibfnamefont {E.}~\bibnamefont
  {Schroten}}, \bibinfo {author} {\bibfnamefont {A.}~\bibnamefont {Goossens}},
  \ and\ \bibinfo {author} {\bibfnamefont {J.}~\bibnamefont {Schoonman}},\
  }\href {\doibase 10.1063/1.366881} {\bibfield  {journal} {\bibinfo  {journal}
  {Journal of Applied Physics}\ }\textbf {\bibinfo {volume} {83}},\ \bibinfo
  {pages} {1660} (\bibinfo {year} {1998})}\BibitemShut {NoStop}%
\bibitem [{\citenamefont {Zunger}(2003)}]{Zunger2003}%
  \BibitemOpen
  \bibfield  {author} {\bibinfo {author} {\bibfnamefont {A.}~\bibnamefont
  {Zunger}},\ }\href {\doibase 10.1063/1.1584074} {\bibfield  {journal}
  {\bibinfo  {journal} {Applied Physics Letters}\ }\textbf {\bibinfo {volume}
  {83}},\ \bibinfo {pages} {57} (\bibinfo {year} {2003})}\BibitemShut {NoStop}%
\bibitem [{\citenamefont {Goyal}\ \emph {et~al.}(2020)\citenamefont {Goyal},
  \citenamefont {Gorai}, \citenamefont {Anand}, \citenamefont {Toberer},
  \citenamefont {Snyder},\ and\ \citenamefont {Stevanovi{\'{c}}}}]{Goyal2020}%
  \BibitemOpen
  \bibfield  {author} {\bibinfo {author} {\bibfnamefont {A.}~\bibnamefont
  {Goyal}}, \bibinfo {author} {\bibfnamefont {P.}~\bibnamefont {Gorai}},
  \bibinfo {author} {\bibfnamefont {S.}~\bibnamefont {Anand}}, \bibinfo
  {author} {\bibfnamefont {E.~S.}\ \bibnamefont {Toberer}}, \bibinfo {author}
  {\bibfnamefont {G.~J.}\ \bibnamefont {Snyder}}, \ and\ \bibinfo {author}
  {\bibfnamefont {V.}~\bibnamefont {Stevanovi{\'{c}}}},\ }\href {\doibase
  10.1021/acs.chemmater.9b05126} {\bibfield  {journal} {\bibinfo  {journal}
  {Chemistry of Materials}\ }\textbf {\bibinfo {volume} {32}},\ \bibinfo
  {pages} {4467} (\bibinfo {year} {2020})}\BibitemShut {NoStop}%
\bibitem [{\citenamefont {Gordon}(2000)}]{Gordon2000}%
  \BibitemOpen
  \bibfield  {author} {\bibinfo {author} {\bibfnamefont {R.~G.}\ \bibnamefont
  {Gordon}},\ }\href {\doibase 10.1557/mrs2000.151} {\bibfield  {journal}
  {\bibinfo  {journal} {MRS Bulletin}\ }\textbf {\bibinfo {volume} {25}},\
  \bibinfo {pages} {52} (\bibinfo {year} {2000})}\BibitemShut {NoStop}%
\bibitem [{\citenamefont {Wang}\ \emph {et~al.}(1964)\citenamefont {Wang},
  \citenamefont {Cardona},\ and\ \citenamefont {Fischer}}]{Wang1964}%
  \BibitemOpen
  \bibfield  {author} {\bibinfo {author} {\bibfnamefont {C.~C.}\ \bibnamefont
  {Wang}}, \bibinfo {author} {\bibfnamefont {M.}~\bibnamefont {Cardona}}, \
  and\ \bibinfo {author} {\bibfnamefont {A.~G.}\ \bibnamefont {Fischer}},\
  }\href@noop {} {\bibfield  {journal} {\bibinfo  {journal} {RCA Review}\
  }\textbf {\bibinfo {volume} {25}} (\bibinfo {year} {1964})}\BibitemShut
  {NoStop}%
\bibitem [{\citenamefont {Chu}\ \emph {et~al.}(1971)\citenamefont {Chu},
  \citenamefont {Jackson}, \citenamefont {Hyslop},\ and\ \citenamefont
  {Chu}}]{Chu1971}%
  \BibitemOpen
  \bibfield  {author} {\bibinfo {author} {\bibfnamefont {T.~L.}\ \bibnamefont
  {Chu}}, \bibinfo {author} {\bibfnamefont {J.~M.}\ \bibnamefont {Jackson}},
  \bibinfo {author} {\bibfnamefont {A.~E.}\ \bibnamefont {Hyslop}}, \ and\
  \bibinfo {author} {\bibfnamefont {S.~C.}\ \bibnamefont {Chu}},\ }\href
  {\doibase 10.1063/1.1659614} {\bibfield  {journal} {\bibinfo  {journal}
  {Journal of Applied Physics}\ }\textbf {\bibinfo {volume} {42}},\ \bibinfo
  {pages} {420} (\bibinfo {year} {1971})}\BibitemShut {NoStop}%
\bibitem [{\citenamefont {Kato}\ \emph {et~al.}(1977)\citenamefont {Kato},
  \citenamefont {Kammura}, \citenamefont {Iwami},\ and\ \citenamefont
  {Kawabe}}]{Kato1977}%
  \BibitemOpen
  \bibfield  {author} {\bibinfo {author} {\bibfnamefont {N.}~\bibnamefont
  {Kato}}, \bibinfo {author} {\bibfnamefont {W.}~\bibnamefont {Kammura}},
  \bibinfo {author} {\bibfnamefont {M.}~\bibnamefont {Iwami}}, \ and\ \bibinfo
  {author} {\bibfnamefont {K.}~\bibnamefont {Kawabe}},\ }\href {\doibase
  10.1143/JJAP.16.1623} {\bibfield  {journal} {\bibinfo  {journal} {Japanese
  Journal of Applied Physics}\ }\textbf {\bibinfo {volume} {16}},\ \bibinfo
  {pages} {1623} (\bibinfo {year} {1977})}\BibitemShut {NoStop}%
\bibitem [{\citenamefont {Crovetto}\ \emph {et~al.}(2020)\citenamefont
  {Crovetto}, \citenamefont {Hempel}, \citenamefont {Rusu}, \citenamefont
  {Choubrac}, \citenamefont {Kojda}, \citenamefont {Habicht},\ and\
  \citenamefont {Unold}}]{Crovetto2020d}%
  \BibitemOpen
  \bibfield  {author} {\bibinfo {author} {\bibfnamefont {A.}~\bibnamefont
  {Crovetto}}, \bibinfo {author} {\bibfnamefont {H.}~\bibnamefont {Hempel}},
  \bibinfo {author} {\bibfnamefont {M.}~\bibnamefont {Rusu}}, \bibinfo {author}
  {\bibfnamefont {L.}~\bibnamefont {Choubrac}}, \bibinfo {author}
  {\bibfnamefont {D.}~\bibnamefont {Kojda}}, \bibinfo {author} {\bibfnamefont
  {K.}~\bibnamefont {Habicht}}, \ and\ \bibinfo {author} {\bibfnamefont
  {T.}~\bibnamefont {Unold}},\ }\href {\doibase 10.1021/acsami.0c11040}
  {\bibfield  {journal} {\bibinfo  {journal} {ACS Applied Materials and
  Interfaces}\ }\textbf {\bibinfo {volume} {12}},\ \bibinfo {pages} {48741}
  (\bibinfo {year} {2020})}\BibitemShut {NoStop}%
\bibitem [{\citenamefont {Odawara}\ \emph {et~al.}(2005)\citenamefont
  {Odawara}, \citenamefont {Udagawa},\ and\ \citenamefont
  {Shimaoka}}]{Odawara2005}%
  \BibitemOpen
  \bibfield  {author} {\bibinfo {author} {\bibfnamefont {M.}~\bibnamefont
  {Odawara}}, \bibinfo {author} {\bibfnamefont {T.}~\bibnamefont {Udagawa}}, \
  and\ \bibinfo {author} {\bibfnamefont {G.}~\bibnamefont {Shimaoka}},\ }\href
  {\doibase 10.1143/JJAP.44.681} {\bibfield  {journal} {\bibinfo  {journal}
  {Japanese Journal of Applied Physics}\ }\textbf {\bibinfo {volume} {44}},\
  \bibinfo {pages} {681} (\bibinfo {year} {2005})}\BibitemShut {NoStop}%
\bibitem [{\citenamefont {Goossens}\ \emph {et~al.}(1989)\citenamefont
  {Goossens}, \citenamefont {Kelder},\ and\ \citenamefont
  {Schoonman}}]{Goossens1989}%
  \BibitemOpen
  \bibfield  {author} {\bibinfo {author} {\bibfnamefont {A.}~\bibnamefont
  {Goossens}}, \bibinfo {author} {\bibfnamefont {E.~M.}\ \bibnamefont
  {Kelder}}, \ and\ \bibinfo {author} {\bibfnamefont {J.}~\bibnamefont
  {Schoonman}},\ }\href {\doibase 10.1002/bbpc.19890931012} {\bibfield
  {journal} {\bibinfo  {journal} {Berichte der Bunsengesellschaft f{\"{u}}r
  physikalische Chemie}\ }\textbf {\bibinfo {volume} {93}},\ \bibinfo {pages}
  {1109} (\bibinfo {year} {1989})}\BibitemShut {NoStop}%
\bibitem [{\citenamefont {Yang}\ \emph {et~al.}(2016)\citenamefont {Yang},
  \citenamefont {Knei$\beta$}, \citenamefont {Lorenz},\ and\ \citenamefont
  {Grundmann}}]{Yang2016a}%
  \BibitemOpen
  \bibfield  {author} {\bibinfo {author} {\bibfnamefont {C.}~\bibnamefont
  {Yang}}, \bibinfo {author} {\bibfnamefont {M.}~\bibnamefont {Knei$\beta$}},
  \bibinfo {author} {\bibfnamefont {M.}~\bibnamefont {Lorenz}}, \ and\ \bibinfo
  {author} {\bibfnamefont {M.}~\bibnamefont {Grundmann}},\ }\href {\doibase
  10.1073/pnas.1613643113} {\bibfield  {journal} {\bibinfo  {journal}
  {Proceedings of the National Academy of Sciences}\ }\textbf {\bibinfo
  {volume} {113}},\ \bibinfo {pages} {12929} (\bibinfo {year}
  {2016})}\BibitemShut {NoStop}%
\bibitem [{\citenamefont {Woo}\ \emph {et~al.}(2016)\citenamefont {Woo},
  \citenamefont {Lee},\ and\ \citenamefont {Kovnir}}]{Woo2016}%
  \BibitemOpen
  \bibfield  {author} {\bibinfo {author} {\bibfnamefont {K.}~\bibnamefont
  {Woo}}, \bibinfo {author} {\bibfnamefont {K.}~\bibnamefont {Lee}}, \ and\
  \bibinfo {author} {\bibfnamefont {K.}~\bibnamefont {Kovnir}},\ }\href
  {\doibase 10.1088/2053-1591/3/7/074003} {\bibfield  {journal} {\bibinfo
  {journal} {Materials Research Express}\ }\textbf {\bibinfo {volume} {3}},\
  \bibinfo {pages} {074003} (\bibinfo {year} {2016})}\BibitemShut {NoStop}%
\bibitem [{\citenamefont {Kumashiro}\ \emph {et~al.}(1997)\citenamefont
  {Kumashiro}, \citenamefont {Yokoyama}, \citenamefont {Sakamoto},\ and\
  \citenamefont {Fujita}}]{Kumashiro1997a}%
  \BibitemOpen
  \bibfield  {author} {\bibinfo {author} {\bibfnamefont {Y.}~\bibnamefont
  {Kumashiro}}, \bibinfo {author} {\bibfnamefont {T.}~\bibnamefont {Yokoyama}},
  \bibinfo {author} {\bibfnamefont {T.}~\bibnamefont {Sakamoto}}, \ and\
  \bibinfo {author} {\bibfnamefont {T.}~\bibnamefont {Fujita}},\ }\href
  {\doibase 10.1006/jssc.1997.7454} {\bibfield  {journal} {\bibinfo  {journal}
  {Journal of Solid State Chemistry}\ }\textbf {\bibinfo {volume} {133}},\
  \bibinfo {pages} {269} (\bibinfo {year} {1997})}\BibitemShut {NoStop}%
\bibitem [{\citenamefont {Dalui}\ \emph {et~al.}(2008)\citenamefont {Dalui},
  \citenamefont {Hussain}, \citenamefont {Varma}, \citenamefont {Paramanik},\
  and\ \citenamefont {Pal}}]{Dalui2008}%
  \BibitemOpen
  \bibfield  {author} {\bibinfo {author} {\bibfnamefont {S.}~\bibnamefont
  {Dalui}}, \bibinfo {author} {\bibfnamefont {S.}~\bibnamefont {Hussain}},
  \bibinfo {author} {\bibfnamefont {S.}~\bibnamefont {Varma}}, \bibinfo
  {author} {\bibfnamefont {D.}~\bibnamefont {Paramanik}}, \ and\ \bibinfo
  {author} {\bibfnamefont {A.}~\bibnamefont {Pal}},\ }\href {\doibase
  10.1016/j.tsf.2007.09.047} {\bibfield  {journal} {\bibinfo  {journal} {Thin
  Solid Films}\ }\textbf {\bibinfo {volume} {516}},\ \bibinfo {pages} {4958}
  (\bibinfo {year} {2008})}\BibitemShut {NoStop}%
\bibitem [{\citenamefont {Yoshioka}\ \emph {et~al.}(2008)\citenamefont
  {Yoshioka}, \citenamefont {Oba}, \citenamefont {Huang}, \citenamefont
  {Tanaka}, \citenamefont {Mizoguchi},\ and\ \citenamefont
  {Yamamoto}}]{Yoshioka2008}%
  \BibitemOpen
  \bibfield  {author} {\bibinfo {author} {\bibfnamefont {S.}~\bibnamefont
  {Yoshioka}}, \bibinfo {author} {\bibfnamefont {F.}~\bibnamefont {Oba}},
  \bibinfo {author} {\bibfnamefont {R.}~\bibnamefont {Huang}}, \bibinfo
  {author} {\bibfnamefont {I.}~\bibnamefont {Tanaka}}, \bibinfo {author}
  {\bibfnamefont {T.}~\bibnamefont {Mizoguchi}}, \ and\ \bibinfo {author}
  {\bibfnamefont {T.}~\bibnamefont {Yamamoto}},\ }\href {\doibase
  10.1063/1.2829785} {\bibfield  {journal} {\bibinfo  {journal} {Journal of
  Applied Physics}\ }\textbf {\bibinfo {volume} {103}},\ \bibinfo {pages}
  {014309} (\bibinfo {year} {2008})}\BibitemShut {NoStop}%
\bibitem [{\citenamefont {Bikowski}\ \emph {et~al.}(2015)\citenamefont
  {Bikowski}, \citenamefont {Rengachari}, \citenamefont {Nie}, \citenamefont
  {Wanderka}, \citenamefont {Stender}, \citenamefont {Schmitz},\ and\
  \citenamefont {Ellmer}}]{Bikowski2015}%
  \BibitemOpen
  \bibfield  {author} {\bibinfo {author} {\bibfnamefont {A.}~\bibnamefont
  {Bikowski}}, \bibinfo {author} {\bibfnamefont {M.}~\bibnamefont
  {Rengachari}}, \bibinfo {author} {\bibfnamefont {M.}~\bibnamefont {Nie}},
  \bibinfo {author} {\bibfnamefont {N.}~\bibnamefont {Wanderka}}, \bibinfo
  {author} {\bibfnamefont {P.}~\bibnamefont {Stender}}, \bibinfo {author}
  {\bibfnamefont {G.}~\bibnamefont {Schmitz}}, \ and\ \bibinfo {author}
  {\bibfnamefont {K.}~\bibnamefont {Ellmer}},\ }\href {\doibase
  10.1063/1.4922152} {\bibfield  {journal} {\bibinfo  {journal} {APL
  Materials}\ }\textbf {\bibinfo {volume} {3}},\ \bibinfo {pages} {060701}
  (\bibinfo {year} {2015})}\BibitemShut {NoStop}%
\bibitem [{\citenamefont {Morales-Masis}\ \emph {et~al.}(2017)\citenamefont
  {Morales-Masis}, \citenamefont {{De Wolf}}, \citenamefont {Woods-Robinson},
  \citenamefont {Ager},\ and\ \citenamefont {Ballif}}]{Morales-Masis2017}%
  \BibitemOpen
  \bibfield  {author} {\bibinfo {author} {\bibfnamefont {M.}~\bibnamefont
  {Morales-Masis}}, \bibinfo {author} {\bibfnamefont {S.}~\bibnamefont {{De
  Wolf}}}, \bibinfo {author} {\bibfnamefont {R.}~\bibnamefont
  {Woods-Robinson}}, \bibinfo {author} {\bibfnamefont {J.~W.}\ \bibnamefont
  {Ager}}, \ and\ \bibinfo {author} {\bibfnamefont {C.}~\bibnamefont
  {Ballif}},\ }\href {\doibase 10.1002/aelm.201600529} {\bibfield  {journal}
  {\bibinfo  {journal} {Advanced Electronic Materials}\ }\textbf {\bibinfo
  {volume} {3}},\ \bibinfo {pages} {1600529} (\bibinfo {year}
  {2017})}\BibitemShut {NoStop}%
\bibitem [{\citenamefont {Szyszka}(2008)}]{Szyszka2008}%
  \BibitemOpen
  \bibfield  {author} {\bibinfo {author} {\bibfnamefont {B.}~\bibnamefont
  {Szyszka}},\ }in\ \href@noop {} {\emph {\bibinfo {booktitle} {Transparent
  Conductive Zinc Oxide: Basics and Applications in Thin Film Solar Cells}}},\
  \bibinfo {editor} {edited by\ \bibinfo {editor} {\bibfnamefont
  {K.}~\bibnamefont {Ellmer}}, \bibinfo {editor} {\bibfnamefont
  {A.}~\bibnamefont {Klein}}, \ and\ \bibinfo {editor} {\bibfnamefont
  {B.}~\bibnamefont {Rech}}}\ (\bibinfo  {publisher} {Springer},\ \bibinfo
  {address} {Berlin},\ \bibinfo {year} {2008})\ p.\ \bibinfo {pages}
  {221}\BibitemShut {NoStop}%
\bibitem [{\citenamefont {Seo}\ \emph {et~al.}(2020)\citenamefont {Seo},
  \citenamefont {Cao}, \citenamefont {Xiao}, \citenamefont {Wu}, \citenamefont
  {Zhao}, \citenamefont {ZHANG},\ and\ \citenamefont {Yan}}]{Seo2020}%
  \BibitemOpen
  \bibfield  {author} {\bibinfo {author} {\bibfnamefont {H.-S.}\ \bibnamefont
  {Seo}}, \bibinfo {author} {\bibfnamefont {W.}~\bibnamefont {Cao}}, \bibinfo
  {author} {\bibfnamefont {J.~C.}\ \bibnamefont {Xiao}}, \bibinfo {author}
  {\bibfnamefont {Y.-C.}\ \bibnamefont {Wu}}, \bibinfo {author} {\bibfnamefont
  {B.}~\bibnamefont {Zhao}}, \bibinfo {author} {\bibfnamefont {X.}~\bibnamefont
  {ZHANG}}, \ and\ \bibinfo {author} {\bibfnamefont {X.}~\bibnamefont {Yan}},\
  }\href {\doibase 10.1002/sdtp.13863} {\bibfield  {journal} {\bibinfo
  {journal} {SID Symposium Digest of Technical Papers}\ }\textbf {\bibinfo
  {volume} {51}},\ \bibinfo {pages} {301} (\bibinfo {year} {2020})}\BibitemShut
  {NoStop}%
\bibitem [{\citenamefont {Jia}\ \emph {et~al.}(2011)\citenamefont {Jia},
  \citenamefont {Zhu}, \citenamefont {Jiang}, \citenamefont {Shen},
  \citenamefont {Han},\ and\ \citenamefont {Chen}}]{Jia2011}%
  \BibitemOpen
  \bibfield  {author} {\bibinfo {author} {\bibfnamefont {Z.}~\bibnamefont
  {Jia}}, \bibinfo {author} {\bibfnamefont {J.}~\bibnamefont {Zhu}}, \bibinfo
  {author} {\bibfnamefont {C.}~\bibnamefont {Jiang}}, \bibinfo {author}
  {\bibfnamefont {W.}~\bibnamefont {Shen}}, \bibinfo {author} {\bibfnamefont
  {J.}~\bibnamefont {Han}}, \ and\ \bibinfo {author} {\bibfnamefont
  {R.}~\bibnamefont {Chen}},\ }\href {\doibase 10.1016/j.apsusc.2011.08.067}
  {\bibfield  {journal} {\bibinfo  {journal} {Applied Surface Science}\
  }\textbf {\bibinfo {volume} {258}},\ \bibinfo {pages} {356} (\bibinfo {year}
  {2011})}\BibitemShut {NoStop}%
\bibitem [{\citenamefont {Xia}\ \emph {et~al.}(1993)\citenamefont {Xia},
  \citenamefont {Xia},\ and\ \citenamefont {Ruoff}}]{Xia1993}%
  \BibitemOpen
  \bibfield  {author} {\bibinfo {author} {\bibfnamefont {H.}~\bibnamefont
  {Xia}}, \bibinfo {author} {\bibfnamefont {Q.}~\bibnamefont {Xia}}, \ and\
  \bibinfo {author} {\bibfnamefont {A.~L.}\ \bibnamefont {Ruoff}},\ }\href
  {\doibase 10.1063/1.354817} {\bibfield  {journal} {\bibinfo  {journal}
  {Journal of Applied Physics}\ }\textbf {\bibinfo {volume} {74}},\ \bibinfo
  {pages} {1660} (\bibinfo {year} {1993})}\BibitemShut {NoStop}%
\bibitem [{\citenamefont {Morosin}\ \emph {et~al.}(1986)\citenamefont
  {Morosin}, \citenamefont {Mullendore}, \citenamefont {Emin},\ and\
  \citenamefont {Slack}}]{Morosin2008}%
  \BibitemOpen
  \bibfield  {author} {\bibinfo {author} {\bibfnamefont {B.}~\bibnamefont
  {Morosin}}, \bibinfo {author} {\bibfnamefont {A.~W.}\ \bibnamefont
  {Mullendore}}, \bibinfo {author} {\bibfnamefont {D.}~\bibnamefont {Emin}}, \
  and\ \bibinfo {author} {\bibfnamefont {G.~A.}\ \bibnamefont {Slack}},\ }in\
  \href {\doibase 10.1063/1.35589} {\emph {\bibinfo {booktitle} {AIP Conference
  Proceedings}}},\ Vol.\ \bibinfo {volume} {140}\ (\bibinfo  {publisher}
  {AIP},\ \bibinfo {year} {1986})\ pp.\ \bibinfo {pages} {70--86}\BibitemShut
  {NoStop}%
\bibitem [{\citenamefont {Solozhenko}\ \emph {et~al.}(2014)\citenamefont
  {Solozhenko}, \citenamefont {Kurakevych}, \citenamefont {{Le Godec}},
  \citenamefont {Kurnosov},\ and\ \citenamefont {Oganov}}]{Solozhenko2014}%
  \BibitemOpen
  \bibfield  {author} {\bibinfo {author} {\bibfnamefont {V.~L.}\ \bibnamefont
  {Solozhenko}}, \bibinfo {author} {\bibfnamefont {O.~O.}\ \bibnamefont
  {Kurakevych}}, \bibinfo {author} {\bibfnamefont {Y.}~\bibnamefont {{Le
  Godec}}}, \bibinfo {author} {\bibfnamefont {A.~V.}\ \bibnamefont {Kurnosov}},
  \ and\ \bibinfo {author} {\bibfnamefont {A.~R.}\ \bibnamefont {Oganov}},\
  }\href {\doibase 10.1063/1.4890231} {\bibfield  {journal} {\bibinfo
  {journal} {Journal of Applied Physics}\ }\textbf {\bibinfo {volume} {116}},\
  \bibinfo {pages} {033501} (\bibinfo {year} {2014})}\BibitemShut {NoStop}%
\bibitem [{\citenamefont {Crovetto}\ \emph {et~al.}(2016)\citenamefont
  {Crovetto}, \citenamefont {Ottsen}, \citenamefont {Stamate}, \citenamefont
  {Kj{\ae}r}, \citenamefont {Schou},\ and\ \citenamefont
  {Hansen}}]{Crovetto2016a}%
  \BibitemOpen
  \bibfield  {author} {\bibinfo {author} {\bibfnamefont {A.}~\bibnamefont
  {Crovetto}}, \bibinfo {author} {\bibfnamefont {T.~S.}\ \bibnamefont
  {Ottsen}}, \bibinfo {author} {\bibfnamefont {E.}~\bibnamefont {Stamate}},
  \bibinfo {author} {\bibfnamefont {D.}~\bibnamefont {Kj{\ae}r}}, \bibinfo
  {author} {\bibfnamefont {J.}~\bibnamefont {Schou}}, \ and\ \bibinfo {author}
  {\bibfnamefont {O.}~\bibnamefont {Hansen}},\ }\href {\doibase
  10.1088/0022-3727/49/29/295101} {\bibfield  {journal} {\bibinfo  {journal}
  {Journal of Physics D: Applied Physics}\ }\textbf {\bibinfo {volume} {49}},\
  \bibinfo {pages} {295101} (\bibinfo {year} {2016})}\BibitemShut {NoStop}%
\bibitem [{\citenamefont {Utsumi}\ \emph {et~al.}(2003)\citenamefont {Utsumi},
  \citenamefont {Iigusa}, \citenamefont {Tokumaru}, \citenamefont {Song},\ and\
  \citenamefont {Shigesato}}]{Utsumi2003}%
  \BibitemOpen
  \bibfield  {author} {\bibinfo {author} {\bibfnamefont {K.}~\bibnamefont
  {Utsumi}}, \bibinfo {author} {\bibfnamefont {H.}~\bibnamefont {Iigusa}},
  \bibinfo {author} {\bibfnamefont {R.}~\bibnamefont {Tokumaru}}, \bibinfo
  {author} {\bibfnamefont {P.}~\bibnamefont {Song}}, \ and\ \bibinfo {author}
  {\bibfnamefont {Y.}~\bibnamefont {Shigesato}},\ }\href {\doibase
  10.1016/S0040-6090(03)01167-2} {\bibfield  {journal} {\bibinfo  {journal}
  {Thin Solid Films}\ }\textbf {\bibinfo {volume} {445}},\ \bibinfo {pages}
  {229} (\bibinfo {year} {2003})}\BibitemShut {NoStop}%
\bibitem [{\citenamefont {Nakamura}\ \emph {et~al.}(1990)\citenamefont
  {Nakamura}, \citenamefont {Fujitsuka},\ and\ \citenamefont
  {Kitajima}}]{Nakamura1990}%
  \BibitemOpen
  \bibfield  {author} {\bibinfo {author} {\bibfnamefont {K.}~\bibnamefont
  {Nakamura}}, \bibinfo {author} {\bibfnamefont {M.}~\bibnamefont {Fujitsuka}},
  \ and\ \bibinfo {author} {\bibfnamefont {M.}~\bibnamefont {Kitajima}},\
  }\href {\doibase 10.1103/PhysRevB.41.12260} {\bibfield  {journal} {\bibinfo
  {journal} {Physical Review B}\ }\textbf {\bibinfo {volume} {41}},\ \bibinfo
  {pages} {12260} (\bibinfo {year} {1990})}\BibitemShut {NoStop}%
\bibitem [{\citenamefont {Fasol}\ \emph {et~al.}(1984)\citenamefont {Fasol},
  \citenamefont {Cardona}, \citenamefont {H{\"{o}}nle},\ and\ \citenamefont
  {von Schnering}}]{Fasol1984}%
  \BibitemOpen
  \bibfield  {author} {\bibinfo {author} {\bibfnamefont {G.}~\bibnamefont
  {Fasol}}, \bibinfo {author} {\bibfnamefont {M.}~\bibnamefont {Cardona}},
  \bibinfo {author} {\bibfnamefont {W.}~\bibnamefont {H{\"{o}}nle}}, \ and\
  \bibinfo {author} {\bibfnamefont {H.}~\bibnamefont {von Schnering}},\ }\href
  {\doibase 10.1016/0038-1098(84)90832-9} {\bibfield  {journal} {\bibinfo
  {journal} {Solid State Communications}\ }\textbf {\bibinfo {volume} {52}},\
  \bibinfo {pages} {307} (\bibinfo {year} {1984})}\BibitemShut {NoStop}%
\bibitem [{\citenamefont {Kuhlmann}\ \emph {et~al.}(1994)\citenamefont
  {Kuhlmann}, \citenamefont {Werheit}, \citenamefont {Lundstr{\"{o}}m},\ and\
  \citenamefont {Robers}}]{Kuhlmann1994}%
  \BibitemOpen
  \bibfield  {author} {\bibinfo {author} {\bibfnamefont {U.}~\bibnamefont
  {Kuhlmann}}, \bibinfo {author} {\bibfnamefont {H.}~\bibnamefont {Werheit}},
  \bibinfo {author} {\bibfnamefont {T.}~\bibnamefont {Lundstr{\"{o}}m}}, \ and\
  \bibinfo {author} {\bibfnamefont {W.}~\bibnamefont {Robers}},\ }\href
  {\doibase 10.1016/0022-3697(94)90056-6} {\bibfield  {journal} {\bibinfo
  {journal} {Journal of Physics and Chemistry of Solids}\ }\textbf {\bibinfo
  {volume} {55}},\ \bibinfo {pages} {579} (\bibinfo {year} {1994})}\BibitemShut
  {NoStop}%
\bibitem [{\citenamefont {de~Keijser}\ \emph {et~al.}(1982)\citenamefont
  {de~Keijser}, \citenamefont {Langford}, \citenamefont {Mittemeijer},\ and\
  \citenamefont {Vogels}}]{DeKeijser1982}%
  \BibitemOpen
  \bibfield  {author} {\bibinfo {author} {\bibfnamefont {T.~H.}\ \bibnamefont
  {de~Keijser}}, \bibinfo {author} {\bibfnamefont {J.~I.}\ \bibnamefont
  {Langford}}, \bibinfo {author} {\bibfnamefont {E.~J.}\ \bibnamefont
  {Mittemeijer}}, \ and\ \bibinfo {author} {\bibfnamefont {A.~B.~P.}\
  \bibnamefont {Vogels}},\ }\href {\doibase 10.1107/S0021889882012035}
  {\bibfield  {journal} {\bibinfo  {journal} {Journal of Applied
  Crystallography}\ }\textbf {\bibinfo {volume} {15}},\ \bibinfo {pages} {308}
  (\bibinfo {year} {1982})}\BibitemShut {NoStop}%
\bibitem [{\citenamefont {Padavala}\ \emph {et~al.}(2018)\citenamefont
  {Padavala}, \citenamefont {{Al Atabi}}, \citenamefont {Tengdelius},
  \citenamefont {Lu}, \citenamefont {H{\"{o}}gberg},\ and\ \citenamefont
  {Edgar}}]{Padavala2018}%
  \BibitemOpen
  \bibfield  {author} {\bibinfo {author} {\bibfnamefont {B.}~\bibnamefont
  {Padavala}}, \bibinfo {author} {\bibfnamefont {H.}~\bibnamefont {{Al
  Atabi}}}, \bibinfo {author} {\bibfnamefont {L.}~\bibnamefont {Tengdelius}},
  \bibinfo {author} {\bibfnamefont {J.}~\bibnamefont {Lu}}, \bibinfo {author}
  {\bibfnamefont {H.}~\bibnamefont {H{\"{o}}gberg}}, \ and\ \bibinfo {author}
  {\bibfnamefont {J.}~\bibnamefont {Edgar}},\ }\href {\doibase
  10.1016/j.jcrysgro.2017.11.014} {\bibfield  {journal} {\bibinfo  {journal}
  {Journal of Crystal Growth}\ }\textbf {\bibinfo {volume} {483}},\ \bibinfo
  {pages} {115} (\bibinfo {year} {2018})}\BibitemShut {NoStop}%
\bibitem [{\citenamefont {Ruck}\ \emph {et~al.}(2005)\citenamefont {Ruck},
  \citenamefont {Hoppe}, \citenamefont {Wahl}, \citenamefont {Simon},
  \citenamefont {Wang},\ and\ \citenamefont {Seifert}}]{Ruck2005}%
  \BibitemOpen
  \bibfield  {author} {\bibinfo {author} {\bibfnamefont {M.}~\bibnamefont
  {Ruck}}, \bibinfo {author} {\bibfnamefont {D.}~\bibnamefont {Hoppe}},
  \bibinfo {author} {\bibfnamefont {B.}~\bibnamefont {Wahl}}, \bibinfo {author}
  {\bibfnamefont {P.}~\bibnamefont {Simon}}, \bibinfo {author} {\bibfnamefont
  {Y.}~\bibnamefont {Wang}}, \ and\ \bibinfo {author} {\bibfnamefont
  {G.}~\bibnamefont {Seifert}},\ }\href {\doibase 10.1002/anie.200503017}
  {\bibfield  {journal} {\bibinfo  {journal} {Angewandte Chemie International
  Edition}\ }\textbf {\bibinfo {volume} {44}},\ \bibinfo {pages} {7616}
  (\bibinfo {year} {2005})}\BibitemShut {NoStop}%
\bibitem [{\citenamefont {Hittorf}(1865)}]{Hittorf1865}%
  \BibitemOpen
  \bibfield  {author} {\bibinfo {author} {\bibfnamefont {W.}~\bibnamefont
  {Hittorf}},\ }\href {\doibase 10.1002/andp.18652021002} {\bibfield  {journal}
  {\bibinfo  {journal} {Annalen der Physik und Chemie}\ }\textbf {\bibinfo
  {volume} {202}},\ \bibinfo {pages} {193} (\bibinfo {year}
  {1865})}\BibitemShut {NoStop}%
\bibitem [{\citenamefont {Motojima}\ \emph {et~al.}(1980)\citenamefont
  {Motojima}, \citenamefont {Yokoe},\ and\ \citenamefont
  {Sugiyama}}]{Motojima1980}%
  \BibitemOpen
  \bibfield  {author} {\bibinfo {author} {\bibfnamefont {S.}~\bibnamefont
  {Motojima}}, \bibinfo {author} {\bibfnamefont {S.}~\bibnamefont {Yokoe}}, \
  and\ \bibinfo {author} {\bibfnamefont {K.}~\bibnamefont {Sugiyama}},\ }\href
  {\doibase 10.1016/0022-0248(80)90054-8} {\bibfield  {journal} {\bibinfo
  {journal} {Journal of Crystal Growth}\ }\textbf {\bibinfo {volume} {49}},\
  \bibinfo {pages} {1} (\bibinfo {year} {1980})}\BibitemShut {NoStop}%
\bibitem [{\citenamefont {Schroten}\ \emph {et~al.}(1996)\citenamefont
  {Schroten}, \citenamefont {Goossens},\ and\ \citenamefont
  {Schoonman}}]{Schroten1996}%
  \BibitemOpen
  \bibfield  {author} {\bibinfo {author} {\bibfnamefont {E.}~\bibnamefont
  {Schroten}}, \bibinfo {author} {\bibfnamefont {A.}~\bibnamefont {Goossens}},
  \ and\ \bibinfo {author} {\bibfnamefont {J.}~\bibnamefont {Schoonman}},\
  }\href {\doibase 10.1063/1.361759} {\bibfield  {journal} {\bibinfo  {journal}
  {Journal of Applied Physics}\ }\textbf {\bibinfo {volume} {79}},\ \bibinfo
  {pages} {4465} (\bibinfo {year} {1996})}\BibitemShut {NoStop}%
\bibitem [{\citenamefont {Slack}\ and\ \citenamefont
  {Bartram}(1975)}]{Slack1975}%
  \BibitemOpen
  \bibfield  {author} {\bibinfo {author} {\bibfnamefont {G.~A.}\ \bibnamefont
  {Slack}}\ and\ \bibinfo {author} {\bibfnamefont {S.~F.}\ \bibnamefont
  {Bartram}},\ }\href {\doibase 10.1063/1.321373} {\bibfield  {journal}
  {\bibinfo  {journal} {Journal of Applied Physics}\ }\textbf {\bibinfo
  {volume} {46}},\ \bibinfo {pages} {89} (\bibinfo {year} {1975})}\BibitemShut
  {NoStop}%
\bibitem [{\citenamefont {Davis}\ and\ \citenamefont {Mott}(1970)}]{Davis1970}%
  \BibitemOpen
  \bibfield  {author} {\bibinfo {author} {\bibfnamefont {E.~A.}\ \bibnamefont
  {Davis}}\ and\ \bibinfo {author} {\bibfnamefont {N.~F.}\ \bibnamefont
  {Mott}},\ }\href {\doibase 10.1080/14786437008221061} {\bibfield  {journal}
  {\bibinfo  {journal} {Philosophical Magazine}\ }\textbf {\bibinfo {volume}
  {22}},\ \bibinfo {pages} {0903} (\bibinfo {year} {1970})}\BibitemShut
  {NoStop}%
\bibitem [{\citenamefont {Schroten}\ \emph {et~al.}(1999)\citenamefont
  {Schroten}, \citenamefont {Goossens},\ and\ \citenamefont
  {Schoonman}}]{Schroten1999}%
  \BibitemOpen
  \bibfield  {author} {\bibinfo {author} {\bibfnamefont {E.}~\bibnamefont
  {Schroten}}, \bibinfo {author} {\bibfnamefont {A.}~\bibnamefont {Goossens}},
  \ and\ \bibinfo {author} {\bibfnamefont {J.}~\bibnamefont {Schoonman}},\
  }\href {\doibase 10.1149/1.1391889} {\bibfield  {journal} {\bibinfo
  {journal} {Journal of The Electrochemical Society}\ }\textbf {\bibinfo
  {volume} {146}},\ \bibinfo {pages} {2045} (\bibinfo {year}
  {1999})}\BibitemShut {NoStop}%
\bibitem [{\citenamefont {Jellison}\ \emph {et~al.}(1993)\citenamefont
  {Jellison}, \citenamefont {Chisholm},\ and\ \citenamefont
  {Gorbatkin}}]{Jellison1993a}%
  \BibitemOpen
  \bibfield  {author} {\bibinfo {author} {\bibfnamefont {G.~E.}\ \bibnamefont
  {Jellison}}, \bibinfo {author} {\bibfnamefont {M.~F.}\ \bibnamefont
  {Chisholm}}, \ and\ \bibinfo {author} {\bibfnamefont {S.~M.}\ \bibnamefont
  {Gorbatkin}},\ }\href {\doibase 10.1063/1.109067} {\bibfield  {journal}
  {\bibinfo  {journal} {Applied Physics Letters}\ }\textbf {\bibinfo {volume}
  {62}},\ \bibinfo {pages} {3348} (\bibinfo {year} {1993})}\BibitemShut
  {NoStop}%
\bibitem [{\citenamefont {Bolat}\ and\ \citenamefont
  {Durandurdu}(2021)}]{Bolat2021}%
  \BibitemOpen
  \bibfield  {author} {\bibinfo {author} {\bibfnamefont {S.}~\bibnamefont
  {Bolat}}\ and\ \bibinfo {author} {\bibfnamefont {M.}~\bibnamefont
  {Durandurdu}},\ }\href {\doibase 10.1016/j.jnoncrysol.2021.121006} {\bibfield
   {journal} {\bibinfo  {journal} {Journal of Non-Crystalline Solids}\ }\textbf
  {\bibinfo {volume} {570}},\ \bibinfo {pages} {121006} (\bibinfo {year}
  {2021})}\BibitemShut {NoStop}%
\bibitem [{\citenamefont {Morita}\ and\ \citenamefont
  {Yamamoto}(1975)}]{Morita1975}%
  \BibitemOpen
  \bibfield  {author} {\bibinfo {author} {\bibfnamefont {N.}~\bibnamefont
  {Morita}}\ and\ \bibinfo {author} {\bibfnamefont {A.}~\bibnamefont
  {Yamamoto}},\ }\href {\doibase 10.1143/JJAP.14.825} {\bibfield  {journal}
  {\bibinfo  {journal} {Japanese Journal of Applied Physics}\ }\textbf
  {\bibinfo {volume} {14}},\ \bibinfo {pages} {825} (\bibinfo {year}
  {1975})}\BibitemShut {NoStop}%
\bibitem [{\citenamefont {Wettling}\ and\ \citenamefont
  {Windscheif}(1984)}]{Wettling1984}%
  \BibitemOpen
  \bibfield  {author} {\bibinfo {author} {\bibfnamefont {W.}~\bibnamefont
  {Wettling}}\ and\ \bibinfo {author} {\bibfnamefont {J.}~\bibnamefont
  {Windscheif}},\ }\href {\doibase 10.1016/0038-1098(84)90053-X} {\bibfield
  {journal} {\bibinfo  {journal} {Solid State Communications}\ }\textbf
  {\bibinfo {volume} {50}},\ \bibinfo {pages} {33} (\bibinfo {year}
  {1984})}\BibitemShut {NoStop}%
\bibitem [{\citenamefont {Swanepoel}(1984)}]{Swanepoel1984}%
  \BibitemOpen
  \bibfield  {author} {\bibinfo {author} {\bibfnamefont {R.}~\bibnamefont
  {Swanepoel}},\ }\href {\doibase 10.1088/0022-3735/17/10/023} {\bibfield
  {journal} {\bibinfo  {journal} {Journal of Physics E: Scientific
  Instruments}\ }\textbf {\bibinfo {volume} {17}},\ \bibinfo {pages} {896}
  (\bibinfo {year} {1984})}\BibitemShut {NoStop}%
\bibitem [{\citenamefont {Crovetto}\ \emph {et~al.}(2018)\citenamefont
  {Crovetto}, \citenamefont {Cazzaniga}, \citenamefont {Ettlinger},
  \citenamefont {Schou},\ and\ \citenamefont {Hansen}}]{Crovetto2018a}%
  \BibitemOpen
  \bibfield  {author} {\bibinfo {author} {\bibfnamefont {A.}~\bibnamefont
  {Crovetto}}, \bibinfo {author} {\bibfnamefont {A.}~\bibnamefont {Cazzaniga}},
  \bibinfo {author} {\bibfnamefont {R.~B.}\ \bibnamefont {Ettlinger}}, \bibinfo
  {author} {\bibfnamefont {J.}~\bibnamefont {Schou}}, \ and\ \bibinfo {author}
  {\bibfnamefont {O.}~\bibnamefont {Hansen}},\ }\href {\doibase
  10.1016/j.solmat.2018.08.005} {\bibfield  {journal} {\bibinfo  {journal}
  {Solar Energy Materials and Solar Cells}\ }\textbf {\bibinfo {volume}
  {187}},\ \bibinfo {pages} {233} (\bibinfo {year} {2018})}\BibitemShut
  {NoStop}%
\bibitem [{\citenamefont {Crovetto}\ \emph {et~al.}(2015)\citenamefont
  {Crovetto}, \citenamefont {Cazzaniga}, \citenamefont {Ettlinger},
  \citenamefont {Schou},\ and\ \citenamefont {Hansen}}]{Crovetto2015}%
  \BibitemOpen
  \bibfield  {author} {\bibinfo {author} {\bibfnamefont {A.}~\bibnamefont
  {Crovetto}}, \bibinfo {author} {\bibfnamefont {A.}~\bibnamefont {Cazzaniga}},
  \bibinfo {author} {\bibfnamefont {R.~B.}\ \bibnamefont {Ettlinger}}, \bibinfo
  {author} {\bibfnamefont {J.}~\bibnamefont {Schou}}, \ and\ \bibinfo {author}
  {\bibfnamefont {O.}~\bibnamefont {Hansen}},\ }\href {\doibase
  10.1016/j.tsf.2014.11.075} {\bibfield  {journal} {\bibinfo  {journal} {Thin
  Solid Films}\ }\textbf {\bibinfo {volume} {582}},\ \bibinfo {pages} {203}
  (\bibinfo {year} {2015})}\BibitemShut {NoStop}%
\bibitem [{\citenamefont {Sirota}(1968)}]{Sirota1968}%
  \BibitemOpen
  \bibfield  {author} {\bibinfo {author} {\bibfnamefont {N.~N.}\ \bibnamefont
  {Sirota}},\ }in\ \href {\doibase 10.1016/S0080-8784(08)60343-9} {\emph
  {\bibinfo {booktitle} {Semiconductors and Semimetals, Voume 4}}}\ (\bibinfo
  {publisher} {Elsevier},\ \bibinfo {year} {1968})\ p.\ \bibinfo {pages}
  {156}\BibitemShut {NoStop}%
\bibitem [{\citenamefont {Houk}(1986)}]{Houk2008}%
  \BibitemOpen
  \bibfield  {author} {\bibinfo {author} {\bibfnamefont {R.~S.}\ \bibnamefont
  {Houk}},\ }\href {\doibase 10.1021/ac00292a003} {\bibfield  {journal}
  {\bibinfo  {journal} {Analytical Chemistry}\ }\textbf {\bibinfo {volume}
  {58}},\ \bibinfo {pages} {97A} (\bibinfo {year} {1986})}\BibitemShut
  {NoStop}%
\bibitem [{\citenamefont {Greenwood}\ and\ \citenamefont
  {Earnshaw}(1997)}]{Greenwood1997}%
  \BibitemOpen
  \bibfield  {author} {\bibinfo {author} {\bibfnamefont {N.~N.}\ \bibnamefont
  {Greenwood}}\ and\ \bibinfo {author} {\bibfnamefont {A.}~\bibnamefont
  {Earnshaw}},\ }\href@noop {} {\emph {\bibinfo {title} {{Chemistry of the
  Elements}}}},\ \bibinfo {edition} {2nd}\ ed.\ (\bibinfo  {publisher}
  {Butterworth-Heineman},\ \bibinfo {address} {Oxford},\ \bibinfo {year}
  {1997})\ pp.\ \bibinfo {pages} {140--143}\BibitemShut {NoStop}%
\bibitem [{\citenamefont {Yu}\ \emph {et~al.}(2015)\citenamefont {Yu},
  \citenamefont {Kong}, \citenamefont {Zhuo}, \citenamefont {Li},\ and\
  \citenamefont {Yao}}]{Yu2015a}%
  \BibitemOpen
  \bibfield  {author} {\bibinfo {author} {\bibfnamefont {D.}~\bibnamefont
  {Yu}}, \bibinfo {author} {\bibfnamefont {C.}~\bibnamefont {Kong}}, \bibinfo
  {author} {\bibfnamefont {J.}~\bibnamefont {Zhuo}}, \bibinfo {author}
  {\bibfnamefont {S.}~\bibnamefont {Li}}, \ and\ \bibinfo {author}
  {\bibfnamefont {Q.}~\bibnamefont {Yao}},\ }\href {\doibase
  10.1007/s11431-015-5841-0} {\bibfield  {journal} {\bibinfo  {journal}
  {Science China Technological Sciences}\ }\textbf {\bibinfo {volume} {58}},\
  \bibinfo {pages} {2016} (\bibinfo {year} {2015})}\BibitemShut {NoStop}%
\bibitem [{\citenamefont {Goossens}\ \emph {et~al.}(1991)\citenamefont
  {Goossens}, \citenamefont {Kelder}, \citenamefont {Beeren}, \citenamefont
  {Bartels},\ and\ \citenamefont {Schoonman}}]{Goossens1991}%
  \BibitemOpen
  \bibfield  {author} {\bibinfo {author} {\bibfnamefont {A.}~\bibnamefont
  {Goossens}}, \bibinfo {author} {\bibfnamefont {E.~M.}\ \bibnamefont
  {Kelder}}, \bibinfo {author} {\bibfnamefont {R.~J.~M.}\ \bibnamefont
  {Beeren}}, \bibinfo {author} {\bibfnamefont {C.~J.~G.}\ \bibnamefont
  {Bartels}}, \ and\ \bibinfo {author} {\bibfnamefont {J.}~\bibnamefont
  {Schoonman}},\ }\href {\doibase 10.1002/bbpc.19910950410} {\bibfield
  {journal} {\bibinfo  {journal} {Berichte der Bunsengesellschaft f{\"{u}}r
  physikalische Chemie}\ }\textbf {\bibinfo {volume} {95}},\ \bibinfo {pages}
  {503} (\bibinfo {year} {1991})}\BibitemShut {NoStop}%
\bibitem [{\citenamefont {Spicer}\ \emph {et~al.}(1980)\citenamefont {Spicer},
  \citenamefont {Lindau}, \citenamefont {Skeath}, \citenamefont {Su},\ and\
  \citenamefont {Chye}}]{Spicer1980}%
  \BibitemOpen
  \bibfield  {author} {\bibinfo {author} {\bibfnamefont {W.~E.}\ \bibnamefont
  {Spicer}}, \bibinfo {author} {\bibfnamefont {I.}~\bibnamefont {Lindau}},
  \bibinfo {author} {\bibfnamefont {P.}~\bibnamefont {Skeath}}, \bibinfo
  {author} {\bibfnamefont {C.~Y.}\ \bibnamefont {Su}}, \ and\ \bibinfo {author}
  {\bibfnamefont {P.}~\bibnamefont {Chye}},\ }\href {\doibase
  10.1103/PhysRevLett.44.420} {\bibfield  {journal} {\bibinfo  {journal}
  {Physical Review Letters}\ }\textbf {\bibinfo {volume} {44}},\ \bibinfo
  {pages} {420} (\bibinfo {year} {1980})}\BibitemShut {NoStop}%
\bibitem [{\citenamefont {Woods-Robinson}\ \emph {et~al.}(2020)\citenamefont
  {Woods-Robinson}, \citenamefont {Ablekim}, \citenamefont {Norman},
  \citenamefont {Johnston}, \citenamefont {Persson}, \citenamefont {Reese},
  \citenamefont {Metzger},\ and\ \citenamefont
  {Zakutayev}}]{Woods-Robinson2020}%
  \BibitemOpen
  \bibfield  {author} {\bibinfo {author} {\bibfnamefont {R.}~\bibnamefont
  {Woods-Robinson}}, \bibinfo {author} {\bibfnamefont {T.}~\bibnamefont
  {Ablekim}}, \bibinfo {author} {\bibfnamefont {A.}~\bibnamefont {Norman}},
  \bibinfo {author} {\bibfnamefont {S.}~\bibnamefont {Johnston}}, \bibinfo
  {author} {\bibfnamefont {K.~A.}\ \bibnamefont {Persson}}, \bibinfo {author}
  {\bibfnamefont {M.~O.}\ \bibnamefont {Reese}}, \bibinfo {author}
  {\bibfnamefont {W.~K.}\ \bibnamefont {Metzger}}, \ and\ \bibinfo {author}
  {\bibfnamefont {A.}~\bibnamefont {Zakutayev}},\ }\href {\doibase
  10.1021/acsaem.0c00413} {\bibfield  {journal} {\bibinfo  {journal} {ACS
  Applied Energy Materials}\ }\textbf {\bibinfo {volume} {3}},\ \bibinfo
  {pages} {5427} (\bibinfo {year} {2020})}\BibitemShut {NoStop}%
\bibitem [{\citenamefont {Saliba}\ \emph {et~al.}(2018)\citenamefont {Saliba},
  \citenamefont {Correa-Baena}, \citenamefont {Wolff}, \citenamefont
  {Stolterfoht}, \citenamefont {Phung}, \citenamefont {Albrecht}, \citenamefont
  {Neher},\ and\ \citenamefont {Abate}}]{Saliba2018}%
  \BibitemOpen
  \bibfield  {author} {\bibinfo {author} {\bibfnamefont {M.}~\bibnamefont
  {Saliba}}, \bibinfo {author} {\bibfnamefont {J.-P.}\ \bibnamefont
  {Correa-Baena}}, \bibinfo {author} {\bibfnamefont {C.~M.}\ \bibnamefont
  {Wolff}}, \bibinfo {author} {\bibfnamefont {M.}~\bibnamefont {Stolterfoht}},
  \bibinfo {author} {\bibfnamefont {N.}~\bibnamefont {Phung}}, \bibinfo
  {author} {\bibfnamefont {S.}~\bibnamefont {Albrecht}}, \bibinfo {author}
  {\bibfnamefont {D.}~\bibnamefont {Neher}}, \ and\ \bibinfo {author}
  {\bibfnamefont {A.}~\bibnamefont {Abate}},\ }\href {\doibase
  10.1021/acs.chemmater.8b00136} {\bibfield  {journal} {\bibinfo  {journal}
  {Chemistry of Materials}\ }\textbf {\bibinfo {volume} {30}},\ \bibinfo
  {pages} {4193} (\bibinfo {year} {2018})}\BibitemShut {NoStop}%
\bibitem [{\citenamefont {Crovetto}\ \emph {et~al.}(2019)\citenamefont
  {Crovetto}, \citenamefont {Nielsen}, \citenamefont {Pandey}, \citenamefont
  {Watts}, \citenamefont {Labram}, \citenamefont {Geisler}, \citenamefont
  {Stenger}, \citenamefont {Jacobsen}, \citenamefont {Hansen}, \citenamefont
  {Seger}, \citenamefont {Chorkendorff},\ and\ \citenamefont
  {Vesborg}}]{Crovetto2019}%
  \BibitemOpen
  \bibfield  {author} {\bibinfo {author} {\bibfnamefont {A.}~\bibnamefont
  {Crovetto}}, \bibinfo {author} {\bibfnamefont {R.}~\bibnamefont {Nielsen}},
  \bibinfo {author} {\bibfnamefont {M.}~\bibnamefont {Pandey}}, \bibinfo
  {author} {\bibfnamefont {L.}~\bibnamefont {Watts}}, \bibinfo {author}
  {\bibfnamefont {J.~G.}\ \bibnamefont {Labram}}, \bibinfo {author}
  {\bibfnamefont {M.}~\bibnamefont {Geisler}}, \bibinfo {author} {\bibfnamefont
  {N.}~\bibnamefont {Stenger}}, \bibinfo {author} {\bibfnamefont {K.~W.}\
  \bibnamefont {Jacobsen}}, \bibinfo {author} {\bibfnamefont {O.}~\bibnamefont
  {Hansen}}, \bibinfo {author} {\bibfnamefont {B.}~\bibnamefont {Seger}},
  \bibinfo {author} {\bibfnamefont {I.}~\bibnamefont {Chorkendorff}}, \ and\
  \bibinfo {author} {\bibfnamefont {P.~C.~K.}\ \bibnamefont {Vesborg}},\ }\href
  {\doibase 10.1021/acs.chemmater.9b00478} {\bibfield  {journal} {\bibinfo
  {journal} {Chemistry of Materials}\ }\textbf {\bibinfo {volume} {31}},\
  \bibinfo {pages} {3359} (\bibinfo {year} {2019})}\BibitemShut {NoStop}%
\bibitem [{\citenamefont {Schnepf}\ \emph {et~al.}(2021)\citenamefont
  {Schnepf}, \citenamefont {Crovetto}, \citenamefont {Gorai}, \citenamefont
  {Park}, \citenamefont {Holtz}, \citenamefont {Heinselman}, \citenamefont
  {Bauers}, \citenamefont {Tellekamp}, \citenamefont {Zakutayev}, \citenamefont
  {Greenaway}, \citenamefont {Toberer},\ and\ \citenamefont
  {Tamboli}}]{Schnepf2021}%
  \BibitemOpen
  \bibfield  {author} {\bibinfo {author} {\bibfnamefont {R.~R.}\ \bibnamefont
  {Schnepf}}, \bibinfo {author} {\bibfnamefont {A.}~\bibnamefont {Crovetto}},
  \bibinfo {author} {\bibfnamefont {P.}~\bibnamefont {Gorai}}, \bibinfo
  {author} {\bibfnamefont {A.}~\bibnamefont {Park}}, \bibinfo {author}
  {\bibfnamefont {M.}~\bibnamefont {Holtz}}, \bibinfo {author} {\bibfnamefont
  {K.}~\bibnamefont {Heinselman}}, \bibinfo {author} {\bibfnamefont {S.~R.}\
  \bibnamefont {Bauers}}, \bibinfo {author} {\bibfnamefont {M.~M.}\
  \bibnamefont {Tellekamp}}, \bibinfo {author} {\bibfnamefont {A.}~\bibnamefont
  {Zakutayev}}, \bibinfo {author} {\bibfnamefont {A.~L.}\ \bibnamefont
  {Greenaway}}, \bibinfo {author} {\bibfnamefont {E.}~\bibnamefont {Toberer}},
  \ and\ \bibinfo {author} {\bibfnamefont {A.~C.}\ \bibnamefont {Tamboli}},\
  }\href {\doibase 10.1039/D1TC04695K} {\bibfield  {journal} {\bibinfo
  {journal} {Journal of Materials Chemistry C}\ ,\ \bibinfo {pages}
  {DOI:10.1039/D1TC04695K}} (\bibinfo {year} {2021})}\BibitemShut {NoStop}%
\bibitem [{\citenamefont {Bock}\ and\ \citenamefont
  {Mueller}(1984)}]{Bock1984}%
  \BibitemOpen
  \bibfield  {author} {\bibinfo {author} {\bibfnamefont {H.}~\bibnamefont
  {Bock}}\ and\ \bibinfo {author} {\bibfnamefont {H.}~\bibnamefont {Mueller}},\
  }\href {\doibase 10.1021/ic00193a051} {\bibfield  {journal} {\bibinfo
  {journal} {Inorganic Chemistry}\ }\textbf {\bibinfo {volume} {23}},\ \bibinfo
  {pages} {4365} (\bibinfo {year} {1984})}\BibitemShut {NoStop}%
\end{thebibliography}%


\begin{thebibliography}{71}%
\makeatletter
\providecommand \@ifxundefined [1]{%
 \@ifx{#1\undefined}
}%
\providecommand \@ifnum [1]{%
 \ifnum #1\expandafter \@firstoftwo
 \else \expandafter \@secondoftwo
 \fi
}%
\providecommand \@ifx [1]{%
 \ifx #1\expandafter \@firstoftwo
 \else \expandafter \@secondoftwo
 \fi
}%
\providecommand \natexlab [1]{#1}%
\providecommand \enquote  [1]{``#1''}%
\providecommand \bibnamefont  [1]{#1}%
\providecommand \bibfnamefont [1]{#1}%
\providecommand \citenamefont [1]{#1}%
\providecommand \href@noop [0]{\@secondoftwo}%
\providecommand \href [0]{\begingroup \@sanitize@url \@href}%
\providecommand \@href[1]{\@@startlink{#1}\@@href}%
\providecommand \@@href[1]{\endgroup#1\@@endlink}%
\providecommand \@sanitize@url [0]{\catcode `\\12\catcode `\$12\catcode
  `\&12\catcode `\#12\catcode `\^12\catcode `\_12\catcode `\%12\relax}%
\providecommand \@@startlink[1]{}%
\providecommand \@@endlink[0]{}%
\providecommand \url  [0]{\begingroup\@sanitize@url \@url }%
\providecommand \@url [1]{\endgroup\@href {#1}{\urlprefix }}%
\providecommand \urlprefix  [0]{URL }%
\providecommand \Eprint [0]{\href }%
\providecommand \doibase [0]{http://dx.doi.org/}%
\providecommand \selectlanguage [0]{\@gobble}%
\providecommand \bibinfo  [0]{\@secondoftwo}%
\providecommand \bibfield  [0]{\@secondoftwo}%
\providecommand \translation [1]{[#1]}%
\providecommand \BibitemOpen [0]{}%
\providecommand \bibitemStop [0]{}%
\providecommand \bibitemNoStop [0]{.\EOS\space}%
\providecommand \EOS [0]{\spacefactor3000\relax}%
\providecommand \BibitemShut  [1]{\csname bibitem#1\endcsname}%
\let\auto@bib@innerbib\@empty
\bibitem [{\citenamefont {Nakamura}, \citenamefont {Fujitsuka},\ and\
  \citenamefont {Kitajima}(1990)}]{Nakamura1990SI}%
  \BibitemOpen
  \bibfield  {author} {\bibinfo {author} {\bibfnamefont {K.}~\bibnamefont
  {Nakamura}}, \bibinfo {author} {\bibfnamefont {M.}~\bibnamefont {Fujitsuka}},
  \ and\ \bibinfo {author} {\bibfnamefont {M.}~\bibnamefont {Kitajima}},\
  }\href {\doibase 10.1103/PhysRevB.41.12260} {\bibfield  {journal} {\bibinfo
  {journal} {Physical Review B}\ }\textbf {\bibinfo {volume} {41}},\ \bibinfo
  {pages} {12260} (\bibinfo {year} {1990})}\BibitemShut {NoStop}%
\bibitem [{\citenamefont {Davis}\ and\ \citenamefont {Mott}(1970)}]{Davis1970SI}%
  \BibitemOpen
  \bibfield  {author} {\bibinfo {author} {\bibfnamefont {E.~A.}\ \bibnamefont
  {Davis}}\ and\ \bibinfo {author} {\bibfnamefont {N.~F.}\ \bibnamefont
  {Mott}},\ }\href {\doibase 10.1080/14786437008221061} {\bibfield  {journal}
  {\bibinfo  {journal} {Philosophical Magazine}\ }\textbf {\bibinfo {volume}
  {22}},\ \bibinfo {pages} {0903} (\bibinfo {year} {1970})}\BibitemShut
  {NoStop}%
\bibitem [{\citenamefont {Schroten}, \citenamefont {Goossens},\ and\
  \citenamefont {Schoonman}(1999)}]{Schroten1999SI}%
  \BibitemOpen
  \bibfield  {author} {\bibinfo {author} {\bibfnamefont {E.}~\bibnamefont
  {Schroten}}, \bibinfo {author} {\bibfnamefont {A.}~\bibnamefont {Goossens}},
  \ and\ \bibinfo {author} {\bibfnamefont {J.}~\bibnamefont {Schoonman}},\
  }\href {\doibase 10.1149/1.1391889} {\bibfield  {journal} {\bibinfo
  {journal} {Journal of The Electrochemical Society}\ }\textbf {\bibinfo
  {volume} {146}},\ \bibinfo {pages} {2045} (\bibinfo {year}
  {1999})}\BibitemShut {NoStop}%
\bibitem [{\citenamefont {Koh}\ \emph {et~al.}(1996)\citenamefont {Koh},
  \citenamefont {Lu}, \citenamefont {Wronski}, \citenamefont {Kuang},
  \citenamefont {Collins}, \citenamefont {Tsong},\ and\ \citenamefont
  {Strausser}}]{Koh1996SI}%
  \BibitemOpen
  \bibfield  {author} {\bibinfo {author} {\bibfnamefont {J.}~\bibnamefont
  {Koh}}, \bibinfo {author} {\bibfnamefont {Y.}~\bibnamefont {Lu}}, \bibinfo
  {author} {\bibfnamefont {C.~R.}\ \bibnamefont {Wronski}}, \bibinfo {author}
  {\bibfnamefont {Y.}~\bibnamefont {Kuang}}, \bibinfo {author} {\bibfnamefont
  {R.~W.}\ \bibnamefont {Collins}}, \bibinfo {author} {\bibfnamefont {T.~T.}\
  \bibnamefont {Tsong}}, \ and\ \bibinfo {author} {\bibfnamefont {Y.~E.}\
  \bibnamefont {Strausser}},\ }\href {\doibase 10.1063/1.117397} {\bibfield
  {journal} {\bibinfo  {journal} {Applied Physics Letters}\ }\textbf {\bibinfo
  {volume} {69}},\ \bibinfo {pages} {1297} (\bibinfo {year}
  {1996})}\BibitemShut {NoStop}%
  \bibitem [{\citenamefont {Hempel}\ \emph {et~al.}(2018)\citenamefont {Hempel},
  \citenamefont {Hages}, \citenamefont {Eichberger}, \citenamefont {Repins},\
  and\ \citenamefont {Unold}}]{Hempel2018aSI}%
  \BibitemOpen
  \bibfield  {author} {\bibinfo {author} {\bibfnamefont {H.}~\bibnamefont
  {Hempel}}, \bibinfo {author} {\bibfnamefont {C.~J.}\ \bibnamefont {Hages}},
  \bibinfo {author} {\bibfnamefont {R.}~\bibnamefont {Eichberger}}, \bibinfo
  {author} {\bibfnamefont {I.}~\bibnamefont {Repins}}, \ and\ \bibinfo {author}
  {\bibfnamefont {T.}~\bibnamefont {Unold}},\ }\href {\doibase
  10.1038/s41598-018-32695-6} {\bibfield  {journal} {\bibinfo  {journal}
  {Scientific Reports}\ }\textbf {\bibinfo {volume} {8}},\ \bibinfo {pages}
  {14476} (\bibinfo {year} {2018})}\BibitemShut {NoStop}%
\end{thebibliography}

\newpage
\clearpage


\onecolumngrid

\section*{SUPPORTING INFORMATION}

\vspace{2cm}

\setcounter{page}{1}
\renewcommand*{\thepage}{S\arabic{page}}

\renewcommand{\thefigure}{S1}
\begin{figure}[h!]
\centering%
\includegraphics[width=0.4\textwidth]{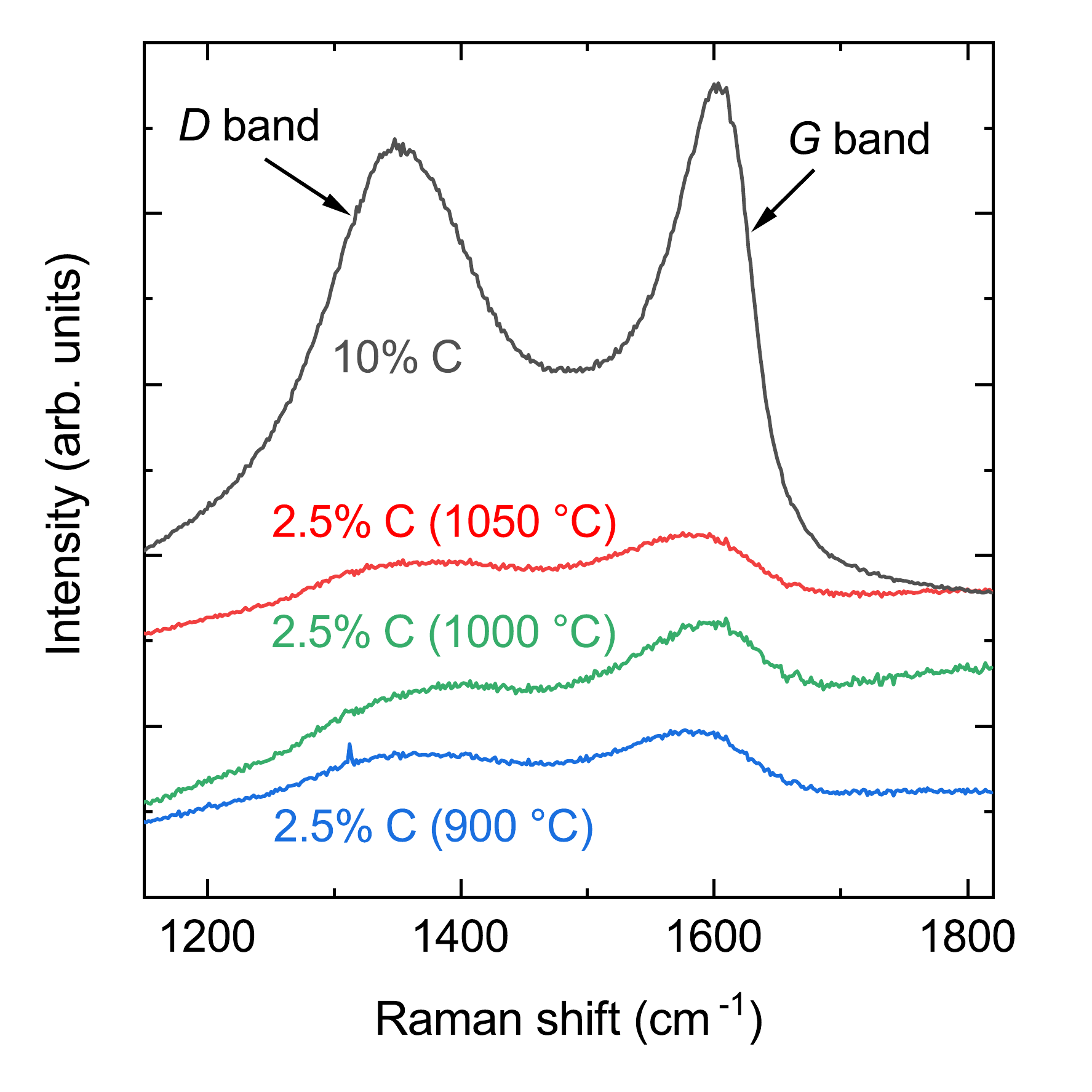}
\caption{Raman spectra of three BP films annealed for 30~min at different temperatures (same samples as in Fig.~\ref{fig:xrd_raman_sem}(b) of the main article). The spectrum of an additional film with higher C content is shown. The two broad peaks present in all films are the $D$ and $G$ bands of disordered carbon.~\cite{Nakamura1990SI}}
\label{fig:SIraman}
\end{figure}

\renewcommand{\thefigure}{S2}
\begin{figure}[h!]
\centering%
\includegraphics[width=0.46\textwidth]{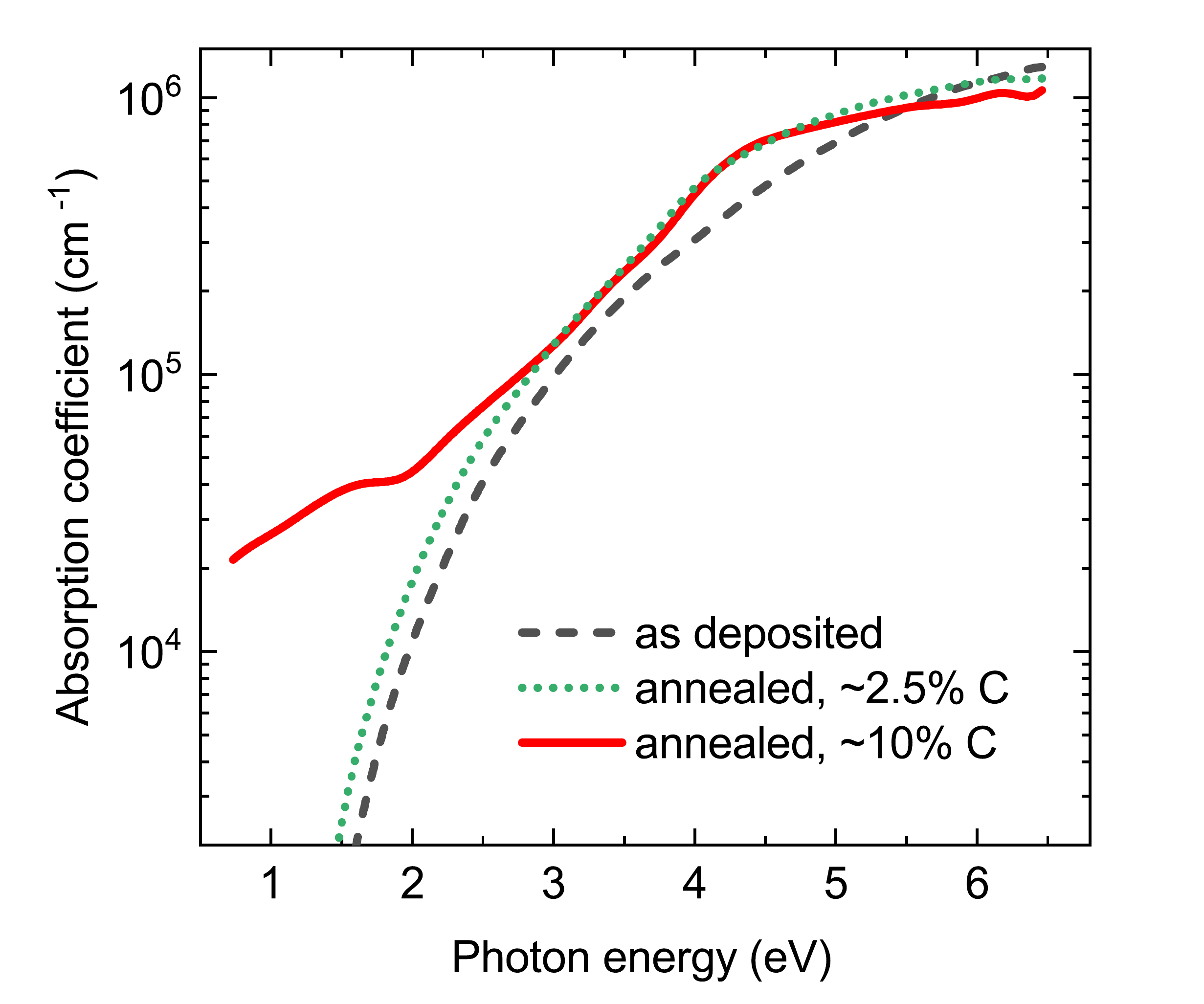}
\caption{Absorption coefficient of an as-deposited BP film, a post-annealed BP film with $\sim$2.5\% C incorporation, and a post-annealed film with $\sim$10\% C incorporation. All three films originate from the same as-deposited film with B/P = 1.3. The main effects of post-annealing for the film with $\sim$2.5\% C are film densification and the appearance of a secondary absorption onset corresponding to the direct band gap. For the post-annealed film with $\sim$10\% C, we assume that significant C precipitation is the cause of stronger absorption below \SI{3}{eV}.}
\label{fig:SIabsorption}
\end{figure}

\renewcommand{\thefigure}{S3}
\begin{figure}[h!]
\centering%
\includegraphics[width=0.5\textwidth]{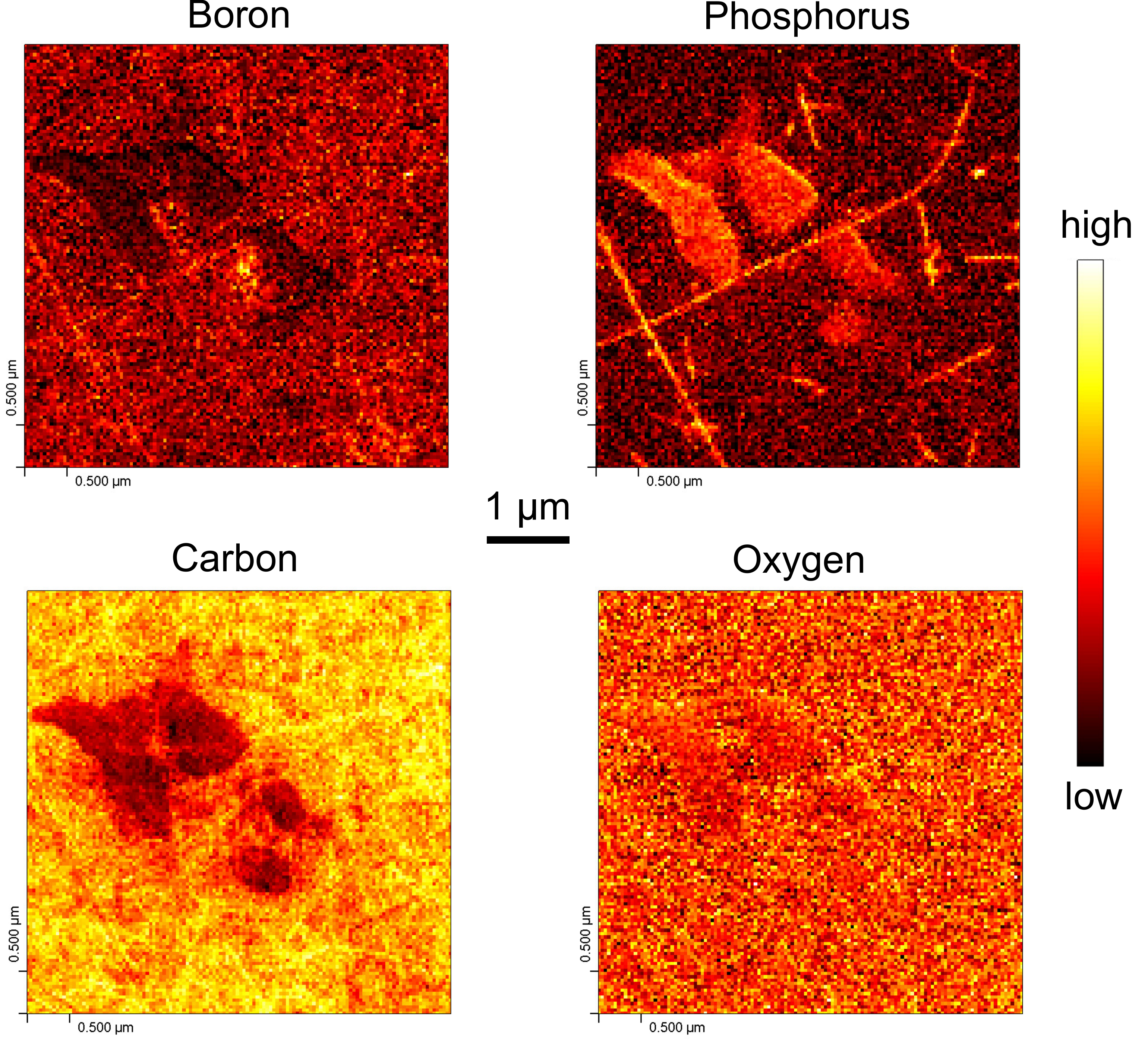}
\caption{AES elemental composition map of a BP film annealed for 24~h with sacrificial P powder. The secondary phases that also visible in SEM images (Fig.~\ref{fig:sem}(b) of the main article) mainly consist of P with some O inclusion. The carbon dopants are homogeneously distributed in the film, but are not incorporated in the secondary phases.}
\label{fig:SIauger_map}
\end{figure}

\renewcommand{\thefigure}{S4}
\begin{figure}[h!]
\centering%
\includegraphics[width=0.9\textwidth]{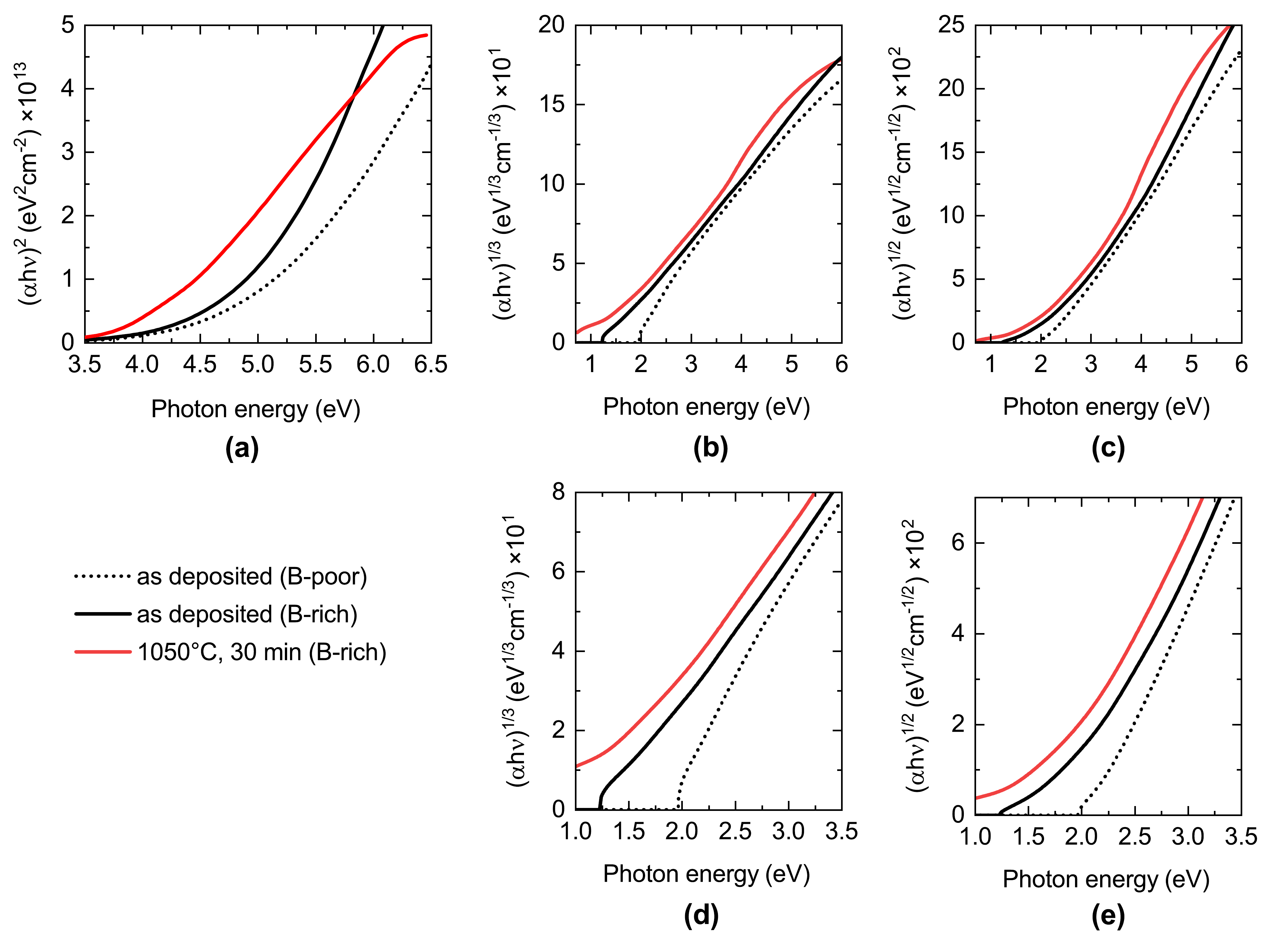}
\caption{Relationships between the absorption coefficient $\alpha$ of BP films and photon energy $h\nu$. The three films that are considered are an as-deposited B-poor film (B/P = 0.6), an as-deposited B-rich film (B/P = 1.3), and a post-annealed film with initial B-rich composition (B/P = 1.3). (a): $(\alpha h \nu)^2$ versus $h \nu$ plot, often used for direct gap crystalline materials. There is a linear region for the post-annealed film only, with estimated direct band gap in the \SI{4.0}\textendash \SI{4.5}{eV} range. (b): $(\alpha h \nu)^{1/3}$ versus $h \nu$ plot. (c): $(\alpha h \nu)^{1/2}$ versus $h \nu$ plot. Both (b) and (c) have physical foundations for amorphous materials, and such materials are often found to obey one or the other.~\cite{Davis1970SI} B-rich B$_\mathrm{x}$P was previously found to obey the $(\alpha h \nu)^{1/3}$ relationship.~\cite{Schroten1999SI} Our B-rich films are also best described by the $(\alpha h \nu)^{1/3}$ relationship for $ h\nu < \SI{4}{eV}$, while the B-poor film is an intermediate case. (d,e): Same plots are (b,c) but closer to the absorption onset.}
\label{fig:SItauc}
\end{figure}

\renewcommand{\thefigure}{S5}
\begin{figure}[h!]
\centering%
\includegraphics[width=0.7\textwidth]{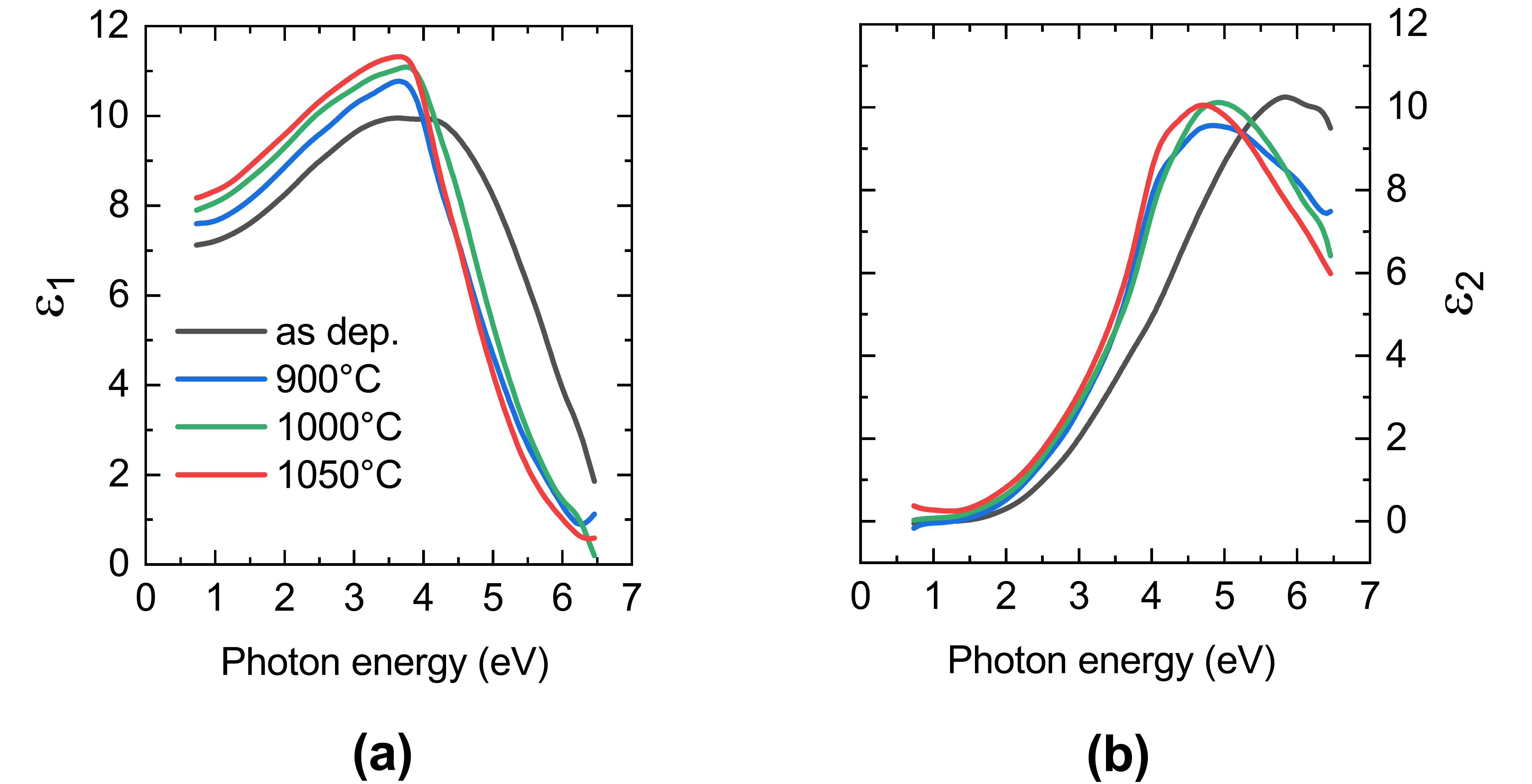}
\caption{Real part ($\epsilon_1$) and imaginary part ($\epsilon_2$) of the dielectric function of the BP films shown in Fig.~\ref{fig:optical}(c,d) in the main article. The high-frequency permittivity $\epsilon_\infty$, estimated by extrapolation of the $\epsilon_1$ spectra to zero photon energy, is between 7 and 8 depending on process conditions.}
\label{fig:SIdielectric_function}
\end{figure}

\renewcommand{\thefigure}{S6}
\begin{figure}[h!]
\centering%
\includegraphics[width=0.55\textwidth]{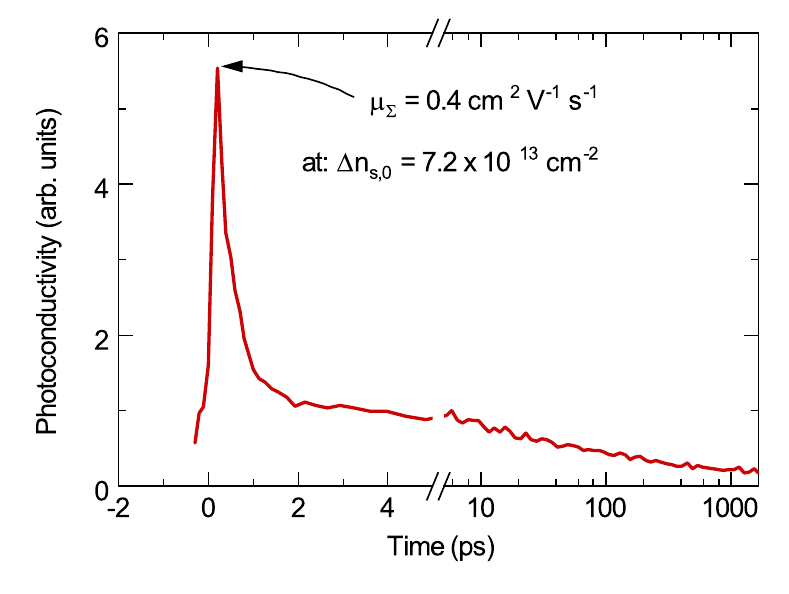}
\caption{Photoconductivity transient after pulsed photoexcitation measured by optical pump terahertz probe spectroscopy. The sum of electron mobility and hole mobility $\mu_\Sigma$ is determined from the initial photoconductivity and the initially photogenerated sheet carrier concentration $\Delta n_{s,0}$ of \SI{7.2 e13}{cm^{-2}}. The pump pulse was the second harmonic of an 150~kHz amplified TiSa femtosecond laser. The terahertz radiation was generated and detected by optical-rectification and electro-optical sampling in ZnTe crystals. For analysis, the thin film approximation was used.~\cite{Hempel2018aSI}}
\label{fig:SITHz}
\end{figure}

\clearpage
\newpage

\begin{center}
\textbf{Conductivity (S/cm)}
\end{center}
\renewcommand{\thetable}{S1}
\begin{table}[h!]
\centering
\setlength{\tabcolsep}{10pt}
\renewcommand{\arraystretch}{1.5}
\begin{tabular}{l c c c}
\hline
 & \SI{900}{\celsius} & \SI{1000}{\celsius} & \SI{1050}{\celsius}  \\
\hline
B/P = 0.92 (C-doped)      & \SI{1.9 e-1} &   \SI{1.6 e-1} &   \SI{-9.9 e-2}{}  \\
B/P = 1.02 (C-doped)      & \SI{2.6 e-2} &   \SI{8.2 e-2} &   \SI{1.8 e-1}{}  \\
B/P = 1.20 (C-doped)      & \SI{7.5 e-2} &   \SI{9.7 e-1} &   \SI{1.9 e+0}{}  \\
B/P = 1.30 (C-doped)      & \SI{2.7 e-2} &   \SI{9.5 e-1} &   \SI{2.1 e+0}{}  \\
B/P = 1.11 (C,Si-doped)  & \SI{3.1 e-3} &   \SI{1.4 e-1} &   \SI{1.5 e+0}{}  \\
\hline
\end{tabular}
\caption{Conductivity of BP films with different B/P ratios and dopants annealed for 30~min at different temperatures. The data in this table is used in Fig.~\ref{fig:electrical}(a) and Fig.~\ref{fig:silicon} of the main article.}
\label{tab:SIconductivity_data}
\end{table}

\vspace{3cm}

\renewcommand{\thetable}{S2}
\begin{table}[h!]
\centering
\setlength{\tabcolsep}{10pt}
\renewcommand{\arraystretch}{1.5}
\begin{tabular}{l c c c}
\hline
 & BP thickness (nm) & Roughness layer thickness (nm) & RMS roughness (nm)  \\
\hline
as-deposited             & 134 &   3.4 &   2.0  \\
\SI{900}{\celsius}       & 123 &   20 &   13  \\
\SI{1000}{\celsius}     & 121 &   16 &   10  \\
\SI{1050}{\celsius}     & 116 &   18 &  12  \\
\hline
\end{tabular}
\caption{Film thickness and roughness layer thickness for various BP films with B/P = 1.30 annealed for 30~min, as determined by ellipsometry. The root mean square (RMS) roughness is estimated using a previously developed formula.~\cite{Koh1996SI}}
\label{tab:SIellipsometry_data}
\end{table}

\vspace{3cm}

\section*{Supplementary references}

\end{document}